\documentclass[a4paper,12pt]{article}

\pdfoutput=1
\usepackage[bbgreekl]{mathbbol}
\usepackage{mathrsfs}
\usepackage{graphicx}
\usepackage{amsmath}
\usepackage{amsfonts}
\usepackage{indentfirst}
\usepackage{amsthm}
\usepackage{amssymb}
\usepackage{xcolor}
\usepackage{geometry}
\usepackage{url}
\usepackage{setspace}
\usepackage{verbatim}
\usepackage{ulem}
\usepackage{multirow}
\usepackage{booktabs}
\usepackage[page,toc,titletoc,title]{appendix}
\usepackage{bm}
\usepackage{cases}
\usepackage{subfigure}
\usepackage[page,toc,titletoc,title]{appendix}
\usepackage{amsmath}

\usepackage[noend]{algpseudocode}
\usepackage{algorithmicx,algorithm}
\usepackage{chngcntr}
\usepackage{ragged2e}
\usepackage{slashed}
\usepackage{caption}

\makeatletter
\renewcommand{\@thesubfigure}{\hskip\subfiglabelskip}
\makeatother

\newcommand{\R}{\mathbb{R}}

\newcommand{\Z}{\mathbb{Z}}

\newcommand{\T}{\mathcal{T}}
\newcommand{\<}{\langle}
\renewcommand{\>}{\rangle}
\newcommand\dd{~{\rm d}}
\newcommand\ham{\mathcal{H}}
\newcommand\rs{r_{\rm s}}

\newcommand\V{\mathcal{V}}
\newcommand\La{\mathcal{L}}

\newcommand\rr{\mathbf{r}}
\newcommand{\st}{\sigma}
\newcommand{\Pa}{\mathcal{P}}
\newcommand{\A}{\mathcal{A}}
\newcommand{\D}{\mathcal{D}}
\newcommand{\I}{\mathcal{I}}
\newcommand{\J}{\mathcal{J}}
\newcommand{\K}{\mathcal{K}}
\newcommand{\Mh}{M_h}
\newcommand{\nFCI}{n_{\rm FCI}}
\newcommand{\hama}{\mathcal{H}_{\alpha}}
\newcommand{\is}{\pmb{i}\pmb{s}}
\newcommand{\jt}{\pmb{j}\pmb{t}}
\newcommand{\xx}{\pmb{x}}
\newcommand{\Ha}{H^{(\alpha)}}

\newcommand{\Hj}{H_{\mathcal{J}}}
\newcommand{\kk}{\mathfrak{K}}

\title{A Finite Element Configuration Interaction Method
\\\vskip 0.1cm
for Wigner Localization
}

\author{
Xue Quan\thanks{
{\tt xuequan@mail.bnu.edu.cn}.
School of Mathematical Sciences, Beijing Normal University, China.
}
~and~ Huajie Chen\thanks{
{\tt chen.huajie@bnu.edu.cn}.
School of Mathematical Sciences, Beijing Normal University, China.
}
}
\date{}

\graphicspath{{images/}}

\begin{document}
\maketitle

\begin{abstract}
The Wigner localization is an electron phase at low densities when the electrons are sharply localized around equilibrium positions.
The simulation of the Wigner localization phenomenon requires careful treatment of the many-body correlations, as the electron-electron interaction dominates the system.
This work proposes a numerical algorithm to study the electron ground states of the Wigner molecules.
The main features of our algorithm are three-fold:
(i) a finite element discretization of the one-body space such that the sharp localization can be captured;
(ii) a good initial state obtained by exploiting the strongly correlated limit;
and (iii) a selected configuration interaction method by choosing the Slater determinants from (stochastic) gradients.
Numerical experiments for some typical one-dimensional quantum wires and two-dimensional circular quantum dots are provided to show the efficiency of our algorithm.
\end{abstract}


\section{Introduction}
\label{sec:introduction}

The Wigner localization is a strong-correlation phenomenon that, at very low densities, the electrons manage to minimize their Coulomb interactions by arranging themselves at specific positions \cite{Auslaender05,Ghosal06,Yannouleas07}.
The Wigner localized states give rise to the so called ``Wigner crystals" and ``Wigner molecules”, which are characterized by strong fluctuations of the density and marked peaks of the density-density correlation function, in analogy with conventional solids and molecules.
The Wigner localized systems have received renewed theoretically and experimentally interests, particularly for low dimensional materials such as 1D quantum wires \cite{Chen2003,Grimes1979,Konik2002,Li2016,Malet2012,Postma2001} 
and 2D quantum dots \cite{Andrei1988,Buhmann1991,Deng2016,Goldys1992,Mendez1983,mendl14,Yannouleas07}.
Further potential applications of the Wigner localized systems include the design and manipulation of qubits and quantum computing devices \cite{Deshpande08,Taylor08,Weiss06,Yannouleas07}, and the realization of infrared sensors for controlling the electron filling in semiconductor nano-structures \cite{Ballester10}.

Along with the fundamental and practical interest, there are well-known challenges to study the Wigner localization by theoretical approaches.
A key feature of such systems is that the Coulomb interactions become dominant over the kinetic term, thus the electrons are strongly correlated and sharply localized around equilibrium positions.
Therefore, there are two main difficulties in numerical simulations of the Wigner localization.
First, the electronic structure models based on mean-field approximations, though requiring relative lower computational cost, can not seize the strong many-body effects well.
Second, one may need an efficient numerical discretization to capture the sharp localization of the electron distributions.
The purpose of this work is to construct a numerical algorithm that can handle the above problems.
Our basic idea is to use the finite elements to discretize the one-electron space and design a configuration interaction (CI) method that can select the determinants on the fly.

The CI methods (see the monograph \cite{Helgaker2000}) have been widely used for simulating many-particle systems.
The full CI (FCI) method \cite{Helgaker2000} is a special case which includes all Slater determinants with proper symmetry.
The number of determinants required in the FCI expansion grows exponentially fast with the number of electrons and one-body basis functions.
There are many variations of CI methods to overcome the problem of unaffordable computational cost of FCI.
One successful approach is based on the stochastic methods, such as the FCI quantum Monte Carlo method \cite{Booth2009,Lu2020} and the fast randomized iteration FCI method \cite{Greene2019,Lim2017}.
The former method describes the wavefunction by empirical distribution of a large number of stochastic walkers, and the latter one stochastically impose sparsity to Hamiltonian matrix and solution vector during the iterations of power method. 
Another widely used approach is the selected CI plus perturbation theory method, which solves the many-body problem within a selected set of determinants \cite{Buenker1974,Huron1973}. 
There are further improvements of this type of methods that accelerate the determinant selecting or the perturbation phase, such as the so-called adaptive sampling CI \cite{Tubman2016} and Heat-bath CI \cite{Holmes2016,Sharma2017}. 
There is a recently developed so-called coordinate descent FCI (CD-FCI) method \cite{Li2019,Lu2019}, which applies an adaptive coordinate descent method to update the coefficients of important determinants during the iterations.
The algorithm developed in this paper conceptually follows the idea of CD-FCI method, by exploiting the descent information of current state to select ``important" determinants in the CI calculations.

The efficiency of our determinant selecting process relies heavily on the initial state, especially the determinants involved in the CI calculations at the beginning of the iterations.
In our work, we exploit the ``semi-classical limit" of the many-particle problem to construct the initial state.
The idea is to ignore the kinetic energy and find the configurations that minimize the electron-electron repulsion.
This construction is highly related to the strictly correlated electrons (SCE) theory, the concept of which was first introduced in \cite{seidl1999} and developed in \cite{Buttazzo2012,Cotar2013,Giorgi2009,Malet2012,Seidl2007}.
The SCE theory can capture the features in high correlation regime and the localization of electrons without breaking the spin or any other symmetries.
Therefore, it is natural to apply this type of methods for simulating the Wigner localized systems \cite{mendl14}.

To discretize the one-electron space, we use the finite element methods such that the sharp localization can be depicted by such local basis functions.
The finite element methods have been successfully applied to many electronic structure calculations, see e.g. \cite{Bao2012,Chen2014,Gavini2007,Pask2001,Suryanarayana2010,Tsuchida1996} and references cited therein. 
We mention that all the above mentioned literature are for effective one-electron problems, and we refer to \cite{Chen2015} for an endeavour to use finite elements for simulating many-body problems.

The rest of this paper is organized as follows.
In Section \ref{sec:gs}, we briefly review the many-body Schr\"{o}dinger equation and introduce a scaling parameter representing the correlation strength of the system.
In Section \ref{sec:algorithm}, we propose a CI algorithm for the ground state calculations of the Wigner localized systems based on finite element discretizations.
In Section \ref{sec:numerics}, we present the numerical experiments of some typical low dimensional systems with Wigner localization. 
Finally, we give some conclusions in Section \ref{sec:conclusion}.


\section{Many-body Schr\"{o}dinger equation}
\label{sec:gs}
\setcounter{equation}{0}

Let $d\in\{1,2\}$ be the dimension of the system and $N\in\Z_+$ be the number of electrons.
We can restrict the electrons such that they lie in a domain $\Omega\subset\R^d$.
Then the electron state of the system is described by the $N$-electron wavefunction $\Psi = \Psi(\rr_1,\st_1, \cdots,\rr_N,\st_N)$ with $\rr_i\in\Omega$ the spatial coordinate and  $\st_i\in\Z_2:=\{\uparrow,\downarrow\}$ the spin variable of the $i$-th electron.
The wavefunction $\Psi$ should belong to the class
\begin{align}
\label{wavefunctionSpace}
\A := \Big\{ \Psi\in L^2\big( (\Omega\times\mathbb{Z}_2)^N;\mathbb{C}\big) :~
\nabla\Psi\in L^2 ,~ 
\Psi ~\mathrm{is~antisymmetric},~\|\Psi\|_{L^2}=1 \Big\} .
\end{align}
Here ``antisymmetric" means that for any permutation $\Pa$, $\Psi(\rr_{\Pa(1)},\st_{\Pa(1)}, \cdots,\rr_{\Pa(N)},\st_{\Pa(N)}) = (-1)^{\epsilon_\Pa} \Psi(\rr_1,\st_1, \cdots,\rr_N,\st_N)$ with $\epsilon_\Pa$ being the parity of the permutation.

Given an external electric field $v_\mathrm{ext}:\R^d\rightarrow\R$, the Hamiltonian of the system is given by
\begin{eqnarray}
\label{hamiltonian}
\ham = -\frac{1}{2}\sum_{i=1}^N\Delta_{\rr_i} + \sum_{i=1}^{N} v_\mathrm{ext}(\rr_i) + \sum_{1\leq i<j\leq N}v_\mathrm{ee}\big(|\rr_i-\rr_j|\big) 
=: T + V_\mathrm{ext} + V_\mathrm{ee} ,
\end{eqnarray}
where $T$ represents the kinetic part of electrons, $V_\mathrm{ext}$ is the external potential, and $V_\mathrm{ee}$ gives the electron-electron interactions with $v_\mathrm{ee}:\R\rightarrow\R$ the Coulomb repulsion. 
Since we focus on the low-dimensional systems with $d\in\{1,2\}$ in this paper, we will use some ``effective" potential $v_\mathrm{ee}(\cdot)$ for electron-electron interactions (see e.g. \cite{bednarek03,friesecke21}), which is not necessarily the bare Coulomb $1/|\cdot|$ as in $\R^3$. 

The ground state of an $N$-electron system can be found by solving the time-independent Schr$\ddot{\mathrm{o}}$dinger equation
\begin{eqnarray}
\label{Schrodinger}
\ham\Psi_0 = E_0\Psi_0,
\end{eqnarray}
where the ground state energy $E_0$ is the lowest eigenvalue of $\ham$ and the ground state wavefunction $\Psi_0$ is the corresponding eigenfunction.
Note that the ground state solution can also be obtained by minimizing the energy functional
\begin{eqnarray}
\label{variation}
E_0 = \min_{\Psi\in \A} \big\< \Psi \big|\ham\big| \Psi \big\> ,
\end{eqnarray}
where we have used the Dirac's bra-ket \cite{Dirac1939} notation.

The aim of this work is to design an efficient numerical scheme to solve \eqref{Schrodinger} or \eqref{variation} for strongly correlated systems, in which the electron density is low and the Wigner localization phenomenon is significant.
To characterize/control the correlation regime, we put a scaling parameter $\alpha$ in the Hamiltonian
\begin{eqnarray}
\label{hamiltonian_scaled}
\hama := \alpha T + V_\mathrm{ee} + V_\mathrm{ext} .
\end{eqnarray}
At small values of $\alpha$, the Coulomb repulsion dominates over the kinetic energy, and the electrons are strongly correlated.
In contrast, when $\alpha$ is large, the kinetic energy dominates and the electrons behave more like non-interacting particles.
Therefore, we can study the crossover from Fermi liquid to Wigner molecule by varying the parameter $\alpha$.

We mention that it is common to characterize the transition/crossover of Wigner crystallization by a single parameter: the so-called Wigner-Seitz radius \cite{Ashcroft1976}. 
The Wigner-Seitz radius $\rs$ represents the radius of a $d$-dimensional sphere containing on average just one electron, which therefore is frequently described by the ``average" electron density.
The single-electron density $\rho$ corresponding to a many-body wavefunction is given by
\begin{eqnarray}
\label{singledensity}
\rho(\rr)= N \sum_{\st_1,\cdots,\st_N\in\Z_2}\int \big| \Psi(\rr,\st_1,\rr_2,\st_2, \dots,\rr_N,\st_N) \big|^2 {\rm d}\rr_2\dots ~{\rm d}\rr_N .
\end{eqnarray}
Then the Wigner-Seitz radius is defined by
\begin{eqnarray}
\label{rs}
\frac{\pi^{\frac{d}{2}}}{\Gamma(1+\frac{d}{2})} \cdot \rs^d
= \frac{1}{\Bar{\rho}},
\end{eqnarray}
where $\Gamma(\cdot)$ denotes the gamma function and $\Bar{\rho}$ is the average electron density, i.e., the number of electrons per unit length or area.
Clearly, the electron density is large for small $\rs$ and the opposite is true for large $\rs$.
We will show in our numerical experiments (see Section \ref{sec:numerics}) that with a given external potential, varying the scaling parameter $\alpha$ corresponds to accessing different regimes of $\rs$.

Specifically, we will consider the ``semi-classical" limit as $\alpha\rightarrow 0$ in \eqref{hamiltonian_scaled}.
At this limit, the system is governed by the external potential and electron-electron interactions, and the variational principle \eqref{variation} (with $\alpha=0$) should be interpreted as for the $N$-point probability measures
\begin{eqnarray}
\label{variation-semiclassical}
\inf_{|\Psi|^2\text{ is a measurement}} \int_{\Omega^N} \big(V_{\rm ee}+V_{\rm ext}\big) \dd |\Psi|^2.
\end{eqnarray}
By enlarging the space of admissible class to the space of probability measures on $\R^{dN}$, one allows the $N$-point densities $|\Psi|^2$ to concentrate on lower dimensional subsets (see discussions in Section \ref{sec:sce} and numerics in Section \ref{sec:numerics}).
We mention that the systems with such interactions have also been studied in the asymptotic limit as the number of particles goes to infinity (see e.g. \cite{friesecke15,fournais18,serfaty18}).
We finally point out that the semi-classical limit has also been investigated within the framework of density functional theory (DFT), which is highly related to this work.
In particular, the semi-classical limit was derived and analyzed for the Hohenberg-Kohn functional when the single-particle density is fixed (see \cite{friesecke22} for a comprehensive review).


\section{A CI algorithm for Wigner localization}
\label{sec:algorithm}
\setcounter{equation}{0}

In this section, we will propose an algorithm to solve the many-electron Schr\"{o}dinger equation with Hamiltonian $\hama$, particularly for systems in the strong correlation regime when $\alpha$ is small.

In our numerical simulations, the electrons are restricted to a box $\Omega=[-L,L]^d$, with appropriate (Dirichlet) boundary conditions.
The algorithm uses the finite elements for one-electron spatial discretization; designs a good starting state based on the semi-classical limit; and selects the Slater determinants based on the gradient information during the iterations.
We will focus on the linear finite elements throughout this paper, but all our constructions can be generalized to higher-order finite elements without difficulty.

\subsection{Finite element discretizations for many-body wavefunctions}
\label{sec:FCIFEM}

We first construct a basis set for one-electron orbitals.
%
For the spatial coordinate, let $\{\T_h\}$ be a shape regular family of nested conforming meshes over $\Omega$ with size $h$.
Let $S^h(\Omega)$ be the space of piecewise linear and continuous functions on $\Omega$:
\begin{eqnarray*}
S^h(\Omega):=\Big\{ u\in C\big(\bar{\Omega}\big) ~:~ u|_{\tau}\in P_{1,\tau} ,~~ \forall\tau\in\T_h \Big\} ,
\end{eqnarray*}
where $P_{1,\tau}$ is the space of linear functions over $\tau$.
Then we have a corresponding one-electron spatial basic set $\{\phi_j\}_{1\leq j\leq\Mh}$ with $\Mh$ being the dimension of $S^h(\Omega)$.
Here $\phi_j$ is the standard finite element basis function that belongs to $S^h(\Omega)$, which equals 1 at the $j$th node on $\T_h$ and 0 at all other nodes.
For the spin coordinate, we have the corresponding spin function space $\D:={\rm span}\big\{\chi_s\big\}_{s\in\Z_2}$, with the basis functions $\chi_{\uparrow}$ and $\chi_{\downarrow}$ satisfying
\begin{eqnarray*}
\label{spin}
\chi_{\uparrow}(\uparrow)=1, ~~ \chi_{\uparrow}(\downarrow)=0
\qquad {\rm and} \qquad
\chi_{\downarrow}(\uparrow)=0, ~~ \chi_{\downarrow}(\downarrow)=1.
\end{eqnarray*}
Then the one-electron orbitals lie in the product space of $S^h(\Omega)$ and $\D$ as
\begin{eqnarray}
\label{1body}
V^h := S^h(\Omega) \otimes \D
~=~ \mathrm{span}\Big\{\phi_j(\rr)\chi_s(\sigma)~:~ 1\leq j\leq \Mh,~s\in\Z_2 \Big\} .
\end{eqnarray}
We see immediately that the dimension of $V^h$ is $2\Mh$ with a given finite element discretization.
 
We can then construct the space to approximate $N$-electron wavefunction.
Let
\begin{eqnarray*}
\V_{h,N} := \bigwedge_{i=1}^N V^h \subset \A ,
\end{eqnarray*}
where the symbol $\bigwedge$ means the usual tensorial product $\otimes$ with the additional requirement that one only keeps the antisymmetrized products. 
As $\V_{h,N}$ is constructed from the one-electron orbital space $V^h$, we have the following basis functions of $\V_{h,N}$ as
\begin{align}
\label{slaterd}
\nonumber
\Phi_{i_1s_1,\cdots,i_Ns_N}(\rr_1,\sigma_1,\cdots,\rr_N,\sigma_N)
& = \dfrac{1}{\sqrt{N!}}\sum_\Pa(-1)^{\epsilon_\Pa}\phi_{i_{\Pa(1)}}(\rr_1)\chi_{s_{\Pa(1)}}(\sigma_1) \cdots \phi_{i_{\Pa(N)}}(\rr_N)\chi_{s_{\Pa(N)}}(\sigma_N)
\\
& = \dfrac{1}{\sqrt{N!}}
\begin{vmatrix}
\phi_{i_1}(\rr_1)\chi_{s_1}(\sigma_1) &\cdots&\phi_{i_N}(\rr_1)\chi_{s_N}(\sigma_1) \\  \vdots &\ddots &\vdots\\ \phi_{i_1}(\rr_N)\chi_{s_1}(\sigma_N) &\cdots&\phi_{i_N}(\rr_N)\chi_{s_N}(\sigma_N)
\end{vmatrix} ,
\end{align}
where $i_1,\cdots,i_N\in\{1,\cdots,\Mh\}$, $s_1,\cdots,s_N\in\Z_2$, $\Pa$ is arbitrary permutation of $\{1,\cdots,N\}$ and $\epsilon_\Pa$ is the permutation parity of $\Pa$.
In the language of quantum chemistry, a function of the form \eqref{slaterd} is called a Slater determinant, and we will denote it by $\Phi_{\is}$ for simplicity of presentations.
Note that the total number of basis functions for $\V_{h,N}$ is a combinatorial number
\begin{eqnarray*}
{\rm dim}(\V_{h,N}) = \binom{2\Mh}{N} =: \nFCI .
\end{eqnarray*}
In the following, we will denote by $\I$ the index set for the basis functions of $\V_{h,N}$, that is
\begin{eqnarray}
\label{index}
\I := \Big\{\is:~ \Phi_{\is} \text{ is a Slater determinant of the form \eqref{slaterd}} \Big\} 
\qquad{\rm with}\quad |\I|=\nFCI .
\end{eqnarray}

The FCI method \cite{Helgaker2000} approximates the ground state energy $E_0$ of \eqref{variation} by finding the energy minimal in the finite dimensional subspace $\V_{h,N}$:
\begin{eqnarray}
\label{FCI}
E_h = \min_{\Psi_h\in \V_{h,N}} \big\< \Psi_h \big|\hama\big| \Psi_h \big\> 
= \min_{\pmb{c}\in\R^{|\I|}}f(\pmb{c})
\qquad {\rm with} \quad 
f(\pmb{c}) := \frac{\pmb{c}^\top \Ha\pmb{c}}{\pmb{c}^\top S\pmb{c}} ,
\end{eqnarray}
where the matrices $\Ha,S\in\R^{|\I|\times|\I|}$ have elements $\big(\Ha\big)_{\is,\,\jt} = \<\Phi_{\is}|\hama|\Phi_{\jt}\>$ and $S_{\is,\,\jt} = \<\Phi_{\is}|\Phi_{\jt}\>$.
One can equivalently solve the corresponding matrix eigenvalue problem $\Ha \pmb{c} = E_h S\pmb{c}$ for the lowest lying eigenvalue $E_h$ to approximate the ground state energy.

The FCI method is very accurate, whose approximation error only comes from the discretization for one-electron orbitals (the error of the finite element approximations in our case).
Unfortunately, it is unaffordable in practical calculations even for a medium electron number $N$, since the degrees of freedom $\nFCI$ grows too fast. 
The effective reduction of $\nFCI$ without sacrificing the accuracy is a major concern in the FCI theory. 
Since the solution to the FCI wavefunction is generally sparse \cite{Anderson18}, the selected CI method has been proposed to exploit the sparsity, which iteratively solves the variational problem within a selected set of determinants \cite{Buenker1974,Huron1973}. 
At each iteration, the ground state of the Hamiltonian $\Hj$ (with the selected indices in $\J\subset\I$) is solved, according to which the most important determinants outside of the current selected set are added to $\J$. 
The process is repeated until some stopping criteria is reached. 
Throughout the process, the determinants that do not significantly contribute to the overall wavefunction are always not selected, so the approximate solutions to the FCI problem can maintain some sparsity. 
%
We will then construct a selected CI algorithm within the finite element discretizations, by exploiting the feature of strongly correlated systems.

\subsection{Initialization by the strongly correlated limit}
\label{sec:sce}

In our numerical algorithm, we will find the ground state solution by minimizing the energy functional.
During the optimization procedure, we will choose appropriate Slater determinants \eqref{slaterd} and update the $N$-electron wavefunction on the fly.
A good initialization of the selected set and starting wavefunction is crucial for the success of our algorithm.

The standard CI methods (and other wavefunction methods) usually take the Hartree-Fock (HF) approximation \cite{Helgaker2000} as the starting point, which uses a single Slater determinant to approximate the wavefunction.
The HF approximation essentially considers the $\alpha\rightarrow\infty$ limit of \eqref{hamiltonian_scaled} and completely neglects the electron correlation, which will lead to large deviations of the true ground state of Wigner localized states.
We will see from the numerical experiments in Section \ref{sec:numerics} that the HF approximation is an inefficient initialization for systems with low electron density.

In contrast to the HF approximation, we will consider the opposite $\alpha\rightarrow 0$ limit as the starting point, which is more appropriate for strongly correlated systems with Wigner localization \cite{Malet2013,mendl14}.
This semi-classical limit has been studied by the SCE theory \cite{Cotar2013,friesecke22,Giorgi2009,Seidl2007,seidl1999} 
within the framework of density functional theory, and we can exploit essentially the same idea to construct the initial state.
More precisely, to obtain the ground state of  \eqref{hamiltonian_scaled} when $\alpha$ is very small, we can start from the $\alpha=0$ limit \eqref{variation-semiclassical}
and find the electron configurations that can minimize the interactions in the given external electric field
\begin{align}
\label{classic}
\min_{(\rr_1,\cdots,\rr_N)\in\Omega^N} F(\rr_1,\dots,\rr_N) 
\quad {\rm with} \quad
F(\rr_1,\dots,\rr_N) := \sum_{1\leq i<j\leq N} v_{\rm ee}\big(|\rr_i-\rr_j|\big) + \sum_{i=1}^{N} v_{\rm ext}(\rr_i) .
\end{align}
If $(\tilde{\rr}_1,\cdots,\tilde{\rr}_N)$ is an $N$-electron configuration that minimize \eqref{classic}, then we can write an $N$-point distribution $|\tilde{\Psi}|^2$ satisfying 
\begin{eqnarray*}
|\tilde{\Psi}(\rr_1,\dots,\rr_N)|^2 = \dfrac{1}{N!}\sum_{\Pa}\delta(\rr_1-\tilde{\rr}_{\Pa(1)})\times\dots\times\delta(\rr_N-\tilde{\rr}_{\Pa(N)}) ,
\end{eqnarray*}
where the sum over all permutations $\Pa$ is to ensure the symmetry of the wavefunction.
One can easily see that $\int_{\Omega^N} \big(V_{\rm ee}+V_{\rm ext}\big) \dd |\tilde{\Psi}|^2$ equals the minimum of \eqref{classic}.
Note that $\tilde{\Psi}$ gives actually an $N$-point distribution rather than an admissible wavefunction in $\A$ since the Dirac-delta function is included in the expression.
Nevertheless, it is highly related to practical systems with $\alpha>0$, where the ground state wavefunction can be viewed as some smoothing of $\tilde{\Psi}$.

Then our construction of the initial state consists of three steps: 
(i) treat the electrons as classical charged particles that interact with one another by (effective) Coulomb potentials and find the configurations that minimize \eqref{classic}; 
(ii) find all the Slater determinants \eqref{slaterd} (within the given finite element discretization) that are related to the configurations; 
and (iii) solve the eigenvalue problem with the selected Slater determinants.

In the first step, we find the set of all electron configurations that minimize \eqref{classic},
\begin{eqnarray}
\label{U}
U := \Big\{ \big(\rr_1,\cdots,\rr_N\big) \in \Omega^N ~:~ \big(\rr_1,\cdots,\rr_N\big) ~{\rm solves~\eqref{classic}} \Big\} .
\end{eqnarray}
Since the kinetic part starts to play a role when $\alpha\neq 0$, the wavefunction that minimizes the total energy will favor some smooth distribution in the region around $U$.
Therefore, we need to involve the whole set $U$ to obtain a good initial guess.
In practice, we apply the Newton method to obtain a local minimizer of \eqref{classic}, and use sufficiently many starting points for the optimization such that the whole set $U$ can be obtained. 

The second step chooses $N$-electron basis functions from the index set $\I$ for initialization.
The idea is to choose the Slater determinants that are ``most related" to the electron configurations in $U$.
In particular, we choose a subset $\I_{\delta}\subset\I$ such that
\begin{multline}
\label{initial-delta}
\I_{\delta} := \Big\{ \is\in\I ~:~
\text{there exist } (\rr_1,\dots,\rr_N)\in U \text{ and a permutation } \Pa \text{ of }\{1,\cdots,N\}
\\
\text{ such that }
\max_{1\leq k\leq N} \big|\xx_{i_k}-\rr_{i_{\Pa(k)}}\big| \leq \delta
\text{ with } \pmb{i} = (i_1,\cdots,i_N) , \Big\} ,
\qquad
\end{multline}
where $\xx_i$ denotes the spatial coordinates of the $i$-th node of the finite element mesh $\T_h$ and $\delta>0$ is a given parameter controlling the size of the initial basis set.

Finally, we construct a Hamiltonian by using the Slater determinants in $\I_{\delta}$, and solve the eigenvalue problem
\begin{eqnarray}
\label{eigen-initial-CI}
\Ha_{\I_\delta} \pmb{c}_{\I_\delta}=E_{\I_\delta}{S_{\I_\delta}}\pmb{c}_{\I_\delta}
\end{eqnarray}
to obtain the approximate ground sate energy $E_{\I_\delta}$ and the corresponding eigen state $\pmb{c}_{\I_\delta}$ within the initial basis set $\I_\delta$.
Note that this is a significantly smaller matrix eigenvalue problem than that of FCI, as the number of basis functions in this initial set grows only linearly with respect to the electron number $N$.

We write the algorithm for initialization in the following.

\begin{algorithm}[H]
	\caption{~Initialization}
    \label{algorithm-ini}
	\vskip 0.1cm
	\hspace*{0.02in} 
	{\bf Input:} Parameter $\delta>0$.
	\begin{algorithmic}[1]
	    \State Solve \eqref{classic} to obtain $U$.
		\State Select $\I_{\delta}$ based on $U$.
        \State Generate $\Ha_{\I_\delta}$ and $S_{\I_\delta}$ within $\mathcal{I}_{\delta}$.
        \State Solve \eqref{eigen-initial-CI} to obtain $\pmb{c}_{\I_\delta}$.
	\end{algorithmic}
	\hspace*{0.02in} 
	{\bf Output:} Initial set $\I_\delta$ and initial state $\pmb{c}_{\I_\delta}$.
\end{algorithm}

Note that the choice of $\delta$ is critical to the initialization algorithm. 
Large values of $\delta$ will lead to large computational cost but a good initial guess, while small values of $\delta$ will give a cheap initialization but a relative worse initial guess. 
For systems with large particle number, the computational cost (though much cheaper than that of FCI method) grows fast as the parameter $\delta$ increases. 
We need to carefully pick up an appropriate value for $\delta$ such that the initialization is reliable while at the same time the computational cost is under control.

\subsection{Selecting the determinants by stochastic gradient}
\label{sec:CDFCI}

Based on the initial set of Slater determinants generated by Algorithm \ref{algorithm-ini}, we can then design a selecting procedure to add more determinants on the fly according to their ``estimated contributions" to the FCI wave function. 
We mention that the selecting scheme we design here conceptually follows the idea of CD-FCI methods \cite{Li2019,Lu2019}, by exploiting the information of the gradients of the current state.

The goal is to find a subset $\J$ of $\I$ adaptively, such that the ``important" Slater determinants for the ground state are contained in $\J$. 
The FCI variational problem \eqref{FCI} is then approximated by
\begin{eqnarray}
\label{optproblem}
\min_{\pmb{c}|_{\J}\neq 0}f(\pmb{c}) ,
\end{eqnarray}
where $\pmb{c}|_{\J}\neq 0$ means that the vector $\pmb{c}\in\R^{|\I|}$ has zero entries corresponding to the Slater determinants in $\I\backslash\J$.
Note that with the given basis set $\J$, \eqref{optproblem} actually gives a CI approximation of the ground state.

In our algorithm, each step of the iteration consists of three parts:
(i) Update the basis set $\J$ by selecting determinants in $\I\backslash\J$ according to the gradient;
(ii) Update the CI state $\pmb{c}$ based on some ``compressed" gradient and line search;
(iii) Update the gradient.

For part (i), let $\J^{(k)}$ be the basis set at the $k$-th step.
Note that at the beginning of the iteration, i.e. when  $k=0$, $\J^{(0)}=\I_{\delta}$ is initialized by Algorithm \ref{algorithm-ini}.
We denote by $\pmb{c}^{(k)}$ the CI solution of \eqref{optproblem} and $\pmb{g}^{(k)}:=\nabla f(\pmb{c}^{(k)})$ the corresponding gradient.
Both $\pmb{c}^{(k)}$ and $\pmb{g}^{(k)}$ have been either initialized or updated in the previous iterations.
We will update $\kk$ entries of the CI coefficients at each step, where $\kk\in\Z_+$ is a fixed parameter.
The $\kk$ Slater determinants are chosen from those that are ``connected" to a randomly selected subset $\La^{(k)} \subset \J^{(k)}$, that is,
\begin{eqnarray*}
\La^{(k)}_{\rm c} := \big\{\jt :~ \big(\Ha\big)_{\is,\jt} \neq 0,~ \is \in \La^{(k)}\big\} .
\end{eqnarray*}
Then we will compare the entries of $\pmb{g}^{(k)}$ within $\La^{(k)}_{\rm c}$ (written as $\pmb{g}^{(k)}|_{\La^{(k)}_{\rm c}}$), and select $\kk$ determinants corresponding to the $\kk$ largest magnitude of $\pmb{g}^{(k)}|_{\La^{(k)}_{\rm c}}$.
However, since the size of $\La^{(k)}_{\rm c}$ grows exponentially with respect to the particle number $N$, it is prohibitive to use this type of selection directly.
We will therefore first perform a stochastic selection in the connected set $\La^{(k)}_{\rm s}\subset \La^{(k)}_{\rm c}$, such that the size of $\La^{(k)}_{\rm s}$ is a fixed number proportional to $\kk$ (which does not depend on the particle number), and then update the basis set by taking $\kk$ determinants with the largest magnitude in $\pmb{g}^{(k)}|_{\La^{(k)}_{\rm s}}$.
More precisely, the updated set for Slater determinants is
\begin{align}
\label{select}
\nonumber
& \J^{(k+1)} = \J^{(k)} \cup \K^{(k)}
\qquad{\rm with}
\\[1ex]
& ~~ \K^{(k)} := \Big\{ \is\in \La^{(k)}_{\rm s} :~ |\pmb{g}^{(k)}|_{\is} \text{ is among the } \kk \text{ largest magnitude of } \pmb{g}^{(k)}|_{\La^{(k)}_{\rm s}} \Big\} .
\end{align}

For part (ii), we will only update the CI state $\pmb{c}^{(k)}$ by the gradient ``compressed" on the set $\K^{(k)}$.
Particularly, we will calculate
\begin{eqnarray}
\label{wfupdate}
\pmb{c}^{(k+1)} = \pmb{c}^{(k)} + \beta^{(k)} \pmb{g}_{\rm s}^{(k)}
\qquad{\rm with}\quad
\pmb{g}_{\rm s}^{(k)} := \pmb{g}^{(k)}|_{\K^{(k)}}
\end{eqnarray}
where the step size $\beta^{(k)}$ can minimize $f\big(\pmb{c}^{(k)}+\beta \pmb{g}_{\rm s}^{(k)}\big)$ with respect to $\beta$.
Note that the derivative of $f$ with respect to $\beta$ derives a quadratic polynomial in $\beta$ and therefore $\beta^{(k)}$ can be obtained explicitly.
 
For part (iii), we can update the gradient (for future iterations) with a relative small cost instead of evaluating $\nabla f$ at each step.
In particular, we have from the ``compressed" gradient $\pmb{g}_{\rm s}^{(k)}$ in \eqref{wfupdate} that
\begin{align}
\label{gradientaprx}
& \pmb{g}^{(k+1)} = \dfrac{2}{(\pmb{c}^{(k+1)})^\top S \pmb{c}^{(k+1)}}\pmb{b}^{(k+1)} 
- \dfrac{2({\pmb{c}^{(k+1)}})^\top \Ha \pmb{c}^{(k+1)}}{\big(({\pmb{c}^{(k+1)}})^\top S \pmb{c}^{(k+1)}\big)^2}\pmb{d}^{(k+1)}
\qquad{\rm with}
\\[1ex] \nonumber
& \pmb{b}^{(k+1)}:= \Ha\pmb{c}^{(k+1)}= \pmb{b}^{(k)}+\beta^{(k)} \Ha\pmb{g}_s^{(k)}\quad \rm{and}\quad
\pmb{d}^{(k+1)}:= S\pmb{c}^{(k+1)}= \pmb{d}^{(k)}+\beta^{(k)} S\pmb{g}_s^{(k)}.
\end{align}
Note that when $k=0$, we have the initial state $\pmb{c}^{(0)}=\pmb{c}_{\I_\delta}$ from Algorithm \ref{algorithm-ini},
so $\pmb{b}$ and $\pmb{d}$ can be directly evaluated at the beginning of the iteration.
To further reduce the computational cost, we can exploit a compression strategy (proposed in \cite{Lim2017,Lu2019}) that ignores the entries of $\pmb{b}^{(k+1)}$ and $\pmb{d}^{(k+1)}$ with increments smaller than some given tolerance.

We can then summarize the complete CI algorithm as follows.

\begin{algorithm}[H]
	\caption{~ CI algorithm for Wigner localization}
	\label{algorithm-CI}
	\vskip 0.1cm
	\hspace*{0.02in} {\bf Input:} 
	Hamiltonian $\hama$;
	$\delta > 0$ (for initialization); 
	$\kk\in \Z_+$ (for size of $\K^{(k)}$).
	\begin{algorithmic}[1]
    	\State 
    	Set $k=0$.
    	Initialize $\J^{(0)}=\I_\delta$ and $\pmb{c}^{(0)} = \pmb{c}_{\I_\delta}$ by Algorithm 3.1.
		\State 
		Calculate $\pmb{b}^{(0)}=\Ha\pmb{c}^{(0)}$, $\pmb{d}^{(0)}=S\pmb{c}^{(0)}$, and $\pmb{g}^{(0)}$ as \eqref{gradientaprx}.
		\While{not converged}
		\State 
		Select $\K^{(k)}$ and update $\J^{(k+1)}$ according to \eqref{select}.
		\State 
		Update $\pmb{c}^{(k+1)}$ as \eqref{wfupdate}.
		\State 
		Calculate $\pmb{g}^{(k+1)}$ as \eqref{gradientaprx}.
		\State
		$k = k +1$
		\EndWhile
		\State{\bf end while}
	\end{algorithmic}
	\hspace*{0.02in} {\bf Output:}
	Basis set $\J$ and ground state wave function approximation $\pmb{c}|_{\J}$.
\end{algorithm}

To stop the iteration in the above algorithm, different criteria can be used to check the ``convergence".
One can monitor the decay of the Rayleigh quotient $f(\pmb{c})$ and stop the iteration as the accumulated values of step size across a few iterations are small.
One can alternatively monitor the gradient of $f(\pmb{c})$ and stop the iteration when the gradient vanishes.
In our implementations, we check the derivatives in some randomly chosen coordinates (over the set $\I$), and stop when the derivatives vanish across a few iterations.

Finally, we would like to comment on the choice of the parameter $\kk$.
By taking larger values of $\kk$, we add more determinants to the basis set at each step, which will accelerate the convergence of the iteration but raise the computational cost significantly.
This trade off can be balanced by carefully choosing some ``optimal" value of $\kk$.
In practical calculations, there is some critical value such that when $\kk$ exceeds this value, the convergence rate will not become faster with the increase of $\kk$ (see our numerical experiments in Section \ref{sec:numerics}).


\section{Numerical experiments}
\label{sec:numerics}
\setcounter{equation}{0}

In this section, we shall demonstrate the efficiency of our algorithm by simulations of several typical 1D and 2D systems with Wigner localization.
All simulations are performed on a workstation with 16 Intel Xeon W-3275M processors and 1T RAM, by using the Julia \cite{Julia} package PairDensities.jl \cite{Xue}.

\subsection{1D systems}
\label{sec:numerics1d}

We consider an $N$-electron system that lies in $\Omega = [-L,L]$, with $L=5.0$ and a given external potential $v_{\rm{ext}}(x) = \frac{1}{2} x^2$. 
The domain $\Omega$ is partitioned by a finite element mesh with $M_h = 50$ interior nodes.
We will consider different scaling parameters $\alpha$ in \eqref{hamiltonian_scaled} with $\alpha$ = 0.1, 1 and 10 respectively.
To test the numerical errors, we use the FCI ground state solution (that is, solution of \eqref{FCI}) as the reference.

\begin{figure}[htb!]
\centering
\subfigure[]
{
\includegraphics[width=4.9cm]{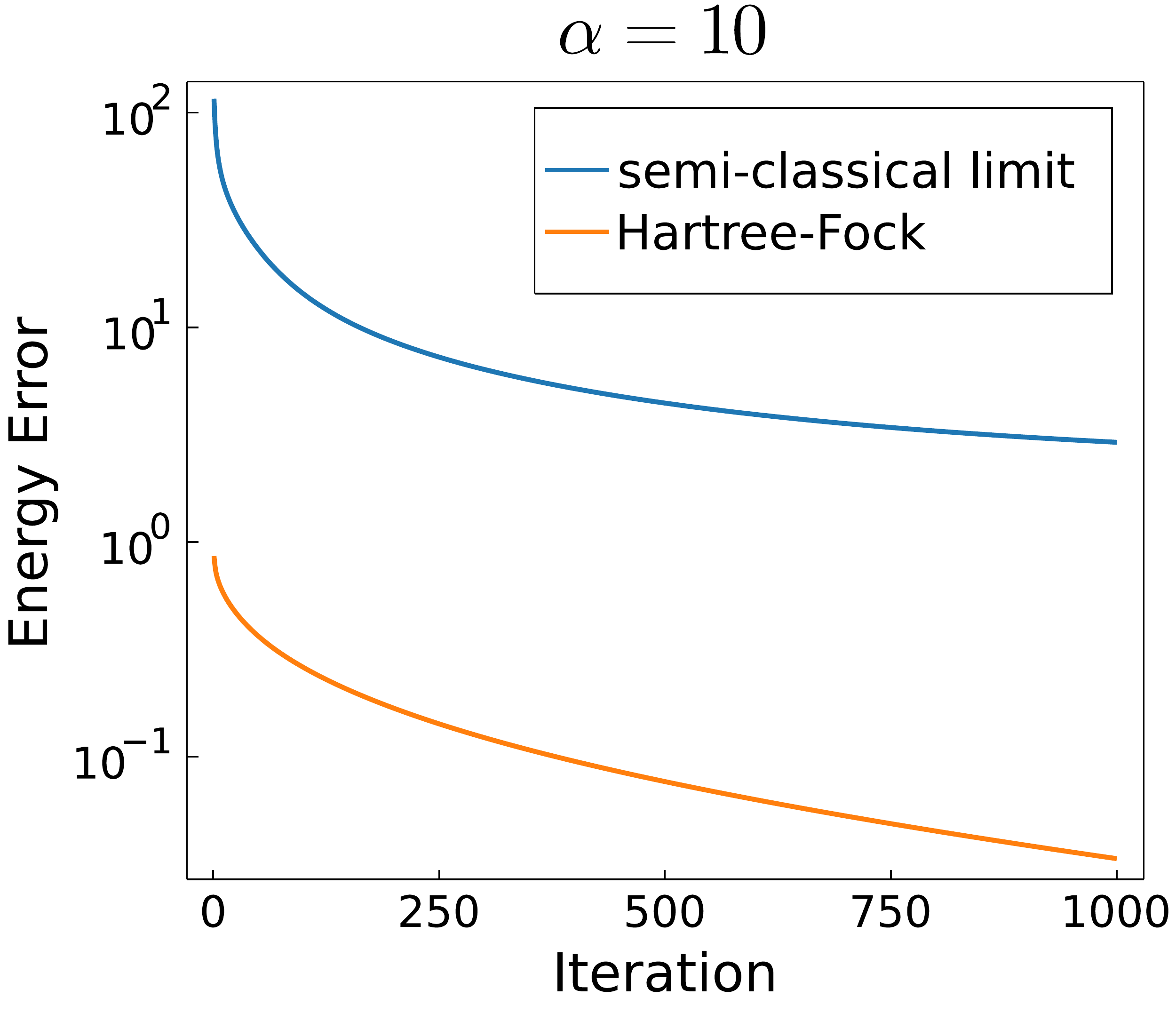}
}
\hskip 0.3cm
\subfigure[]
{
\includegraphics[width=4.9cm]{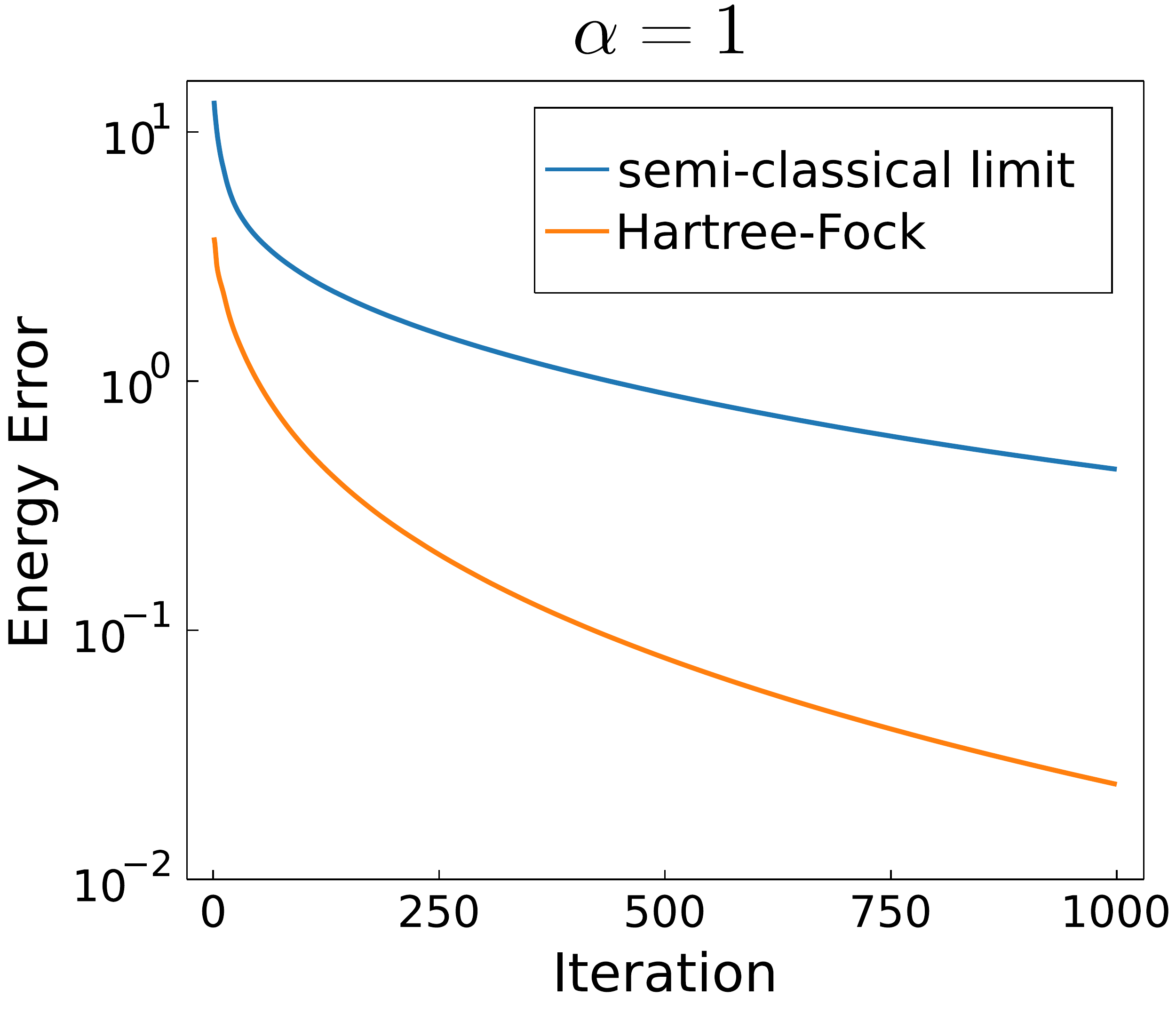}
}
\hskip 0.3cm
\subfigure[]
{
\includegraphics[width=4.9cm]{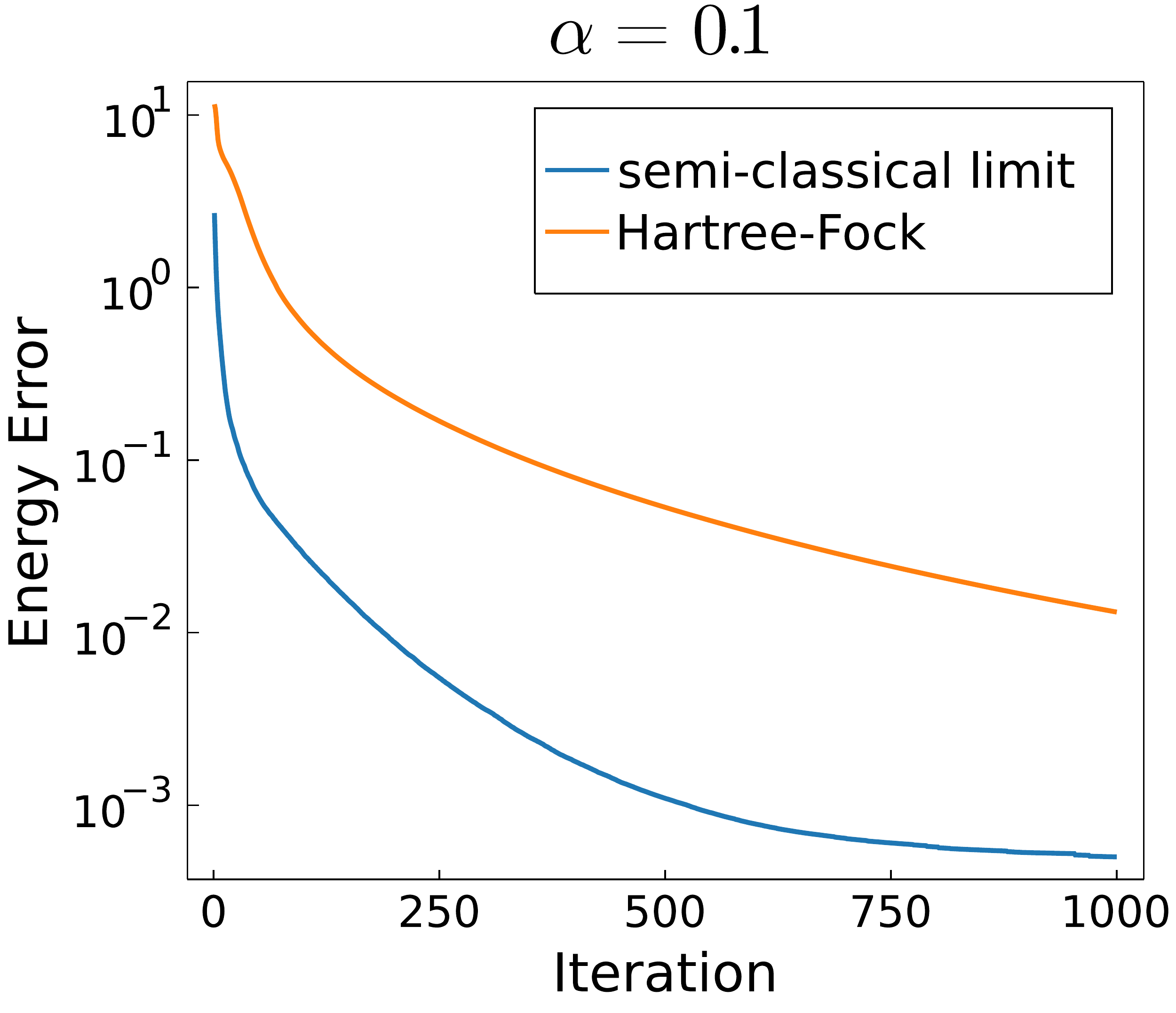}
}
\setlength{\abovecaptionskip}{-4pt}
\caption{Convergence of the ground state energy (for 1D system with $N=4$) with respect to the iterations, where $\alpha = 10$ with $\kk = 5000$, $\alpha = 1$ with $\kk = 1000$, $\alpha = 0.1$ with $\kk = 200$.
}
\label{fig:example1d:err}
\end{figure}

We first test the choice of initial states in Algorithm \ref{algorithm-CI}.
We perform the simulations by using initial state from semi-classical limit (constructed by Algorithm \ref{algorithm-ini}) and that from Hartree-Fock approximation respectively, and compare their convergence in Figure \ref{fig:example1d:err}.
We observe that, when $\alpha$ is large (corresponding to weakly correlated systems), the Hartree-Fock approximation gives a much better approximation; and when $\alpha$ is small (corresponding to strongly correlated systems), the semi-classical limit provides a better initial guess and has a significantly faster convergence rate.
This indicates that our algorithm is potential for systems with strong correlations.

\begin{figure}[htb!]
\centering
\subfigure[]
{
\includegraphics[height=5.5cm]{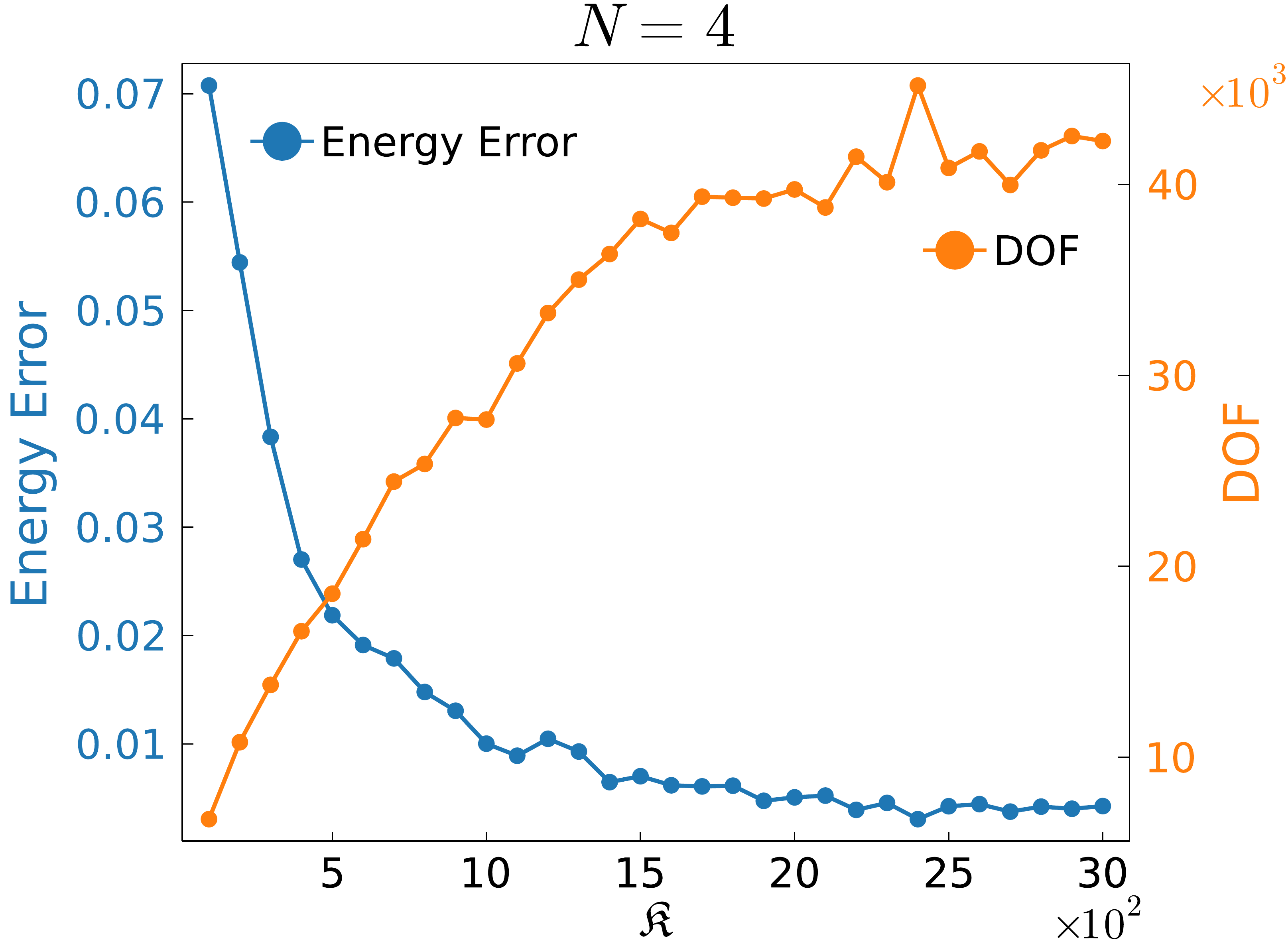}
}
\hskip 0.4cm
\subfigure[]
{
\includegraphics[height=5.5cm]{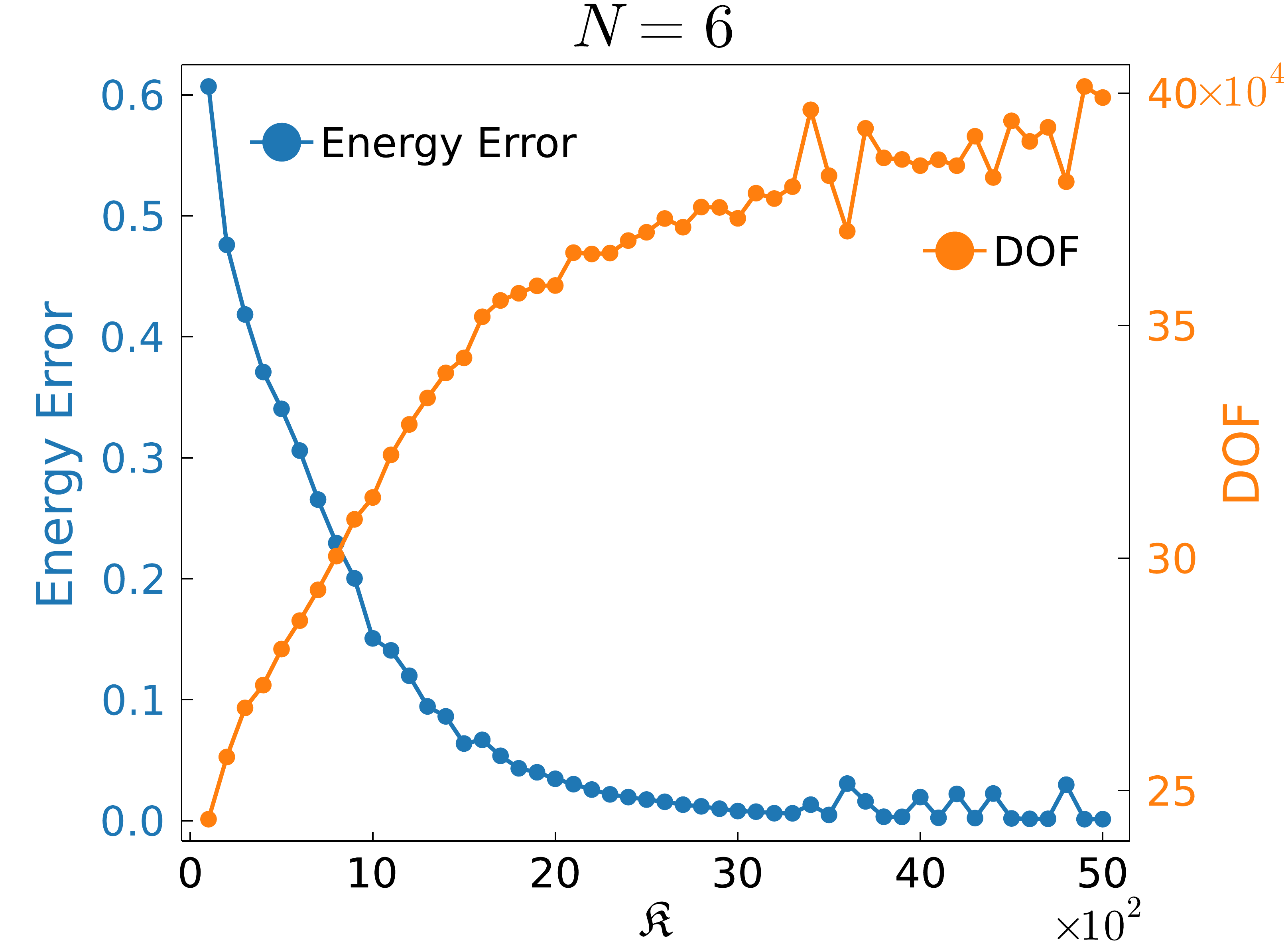}
}
\setlength{\abovecaptionskip}{-4pt}
\caption{The effect of $\kk$ in Algorithm \ref{algorithm-CI}.
The blue lines show the energy errors (with 200 iteration steps) and the orange lines show the growth of the sizes of basis set $\J$.
}
\label{fig:example1d:block}
\end{figure}

We then test the effect of parameter $\kk$ in Algorithm 3.2 by
performing numerical simulations for systems with 4 and 6 electrons respectively, and $\alpha=0.1$.
We run the iterations for 200 steps with different values of $\kk$, and show the numerical errors of ground state approximations and the degrees of freedom (DOF) of basis set $\J$ in Figure \ref{fig:example1d:block}.
We see that for $N=4$, the energy error first decay rapidly when $\kk$ increases from 100 to 1000, then the decay rate becomes much slower and 
reaches a plateau when $\kk \geq 2000$. 
The DOF of basis set $\J$ grows steadily before $\kk = 1500$ and much slower afterwards.
Here, we observe some oscillations of the curves, which can be caused by the stochastic selection of determinants.
As most of the``important" determinants have been selected and added into $\J$, there are very few ``important" ones left outside, which are difficult to be captured by $\La^{(k)}_{\rm s}$.
Similar behaviors of the energy errors and DOF are observed for the $N=6$ case.
We see from the picutures that there could be some ``optimal" choices of $\kk$ for the algorithm, for examples, we can take $\kk$ around 1000 for $N=4$ and $\kk$ around 1500 for $N=6$, such that the accuracy and errors can be well balanced.

\begin{figure}[htb!]
\centering
\subfigure[]
{
\includegraphics[width=7cm]{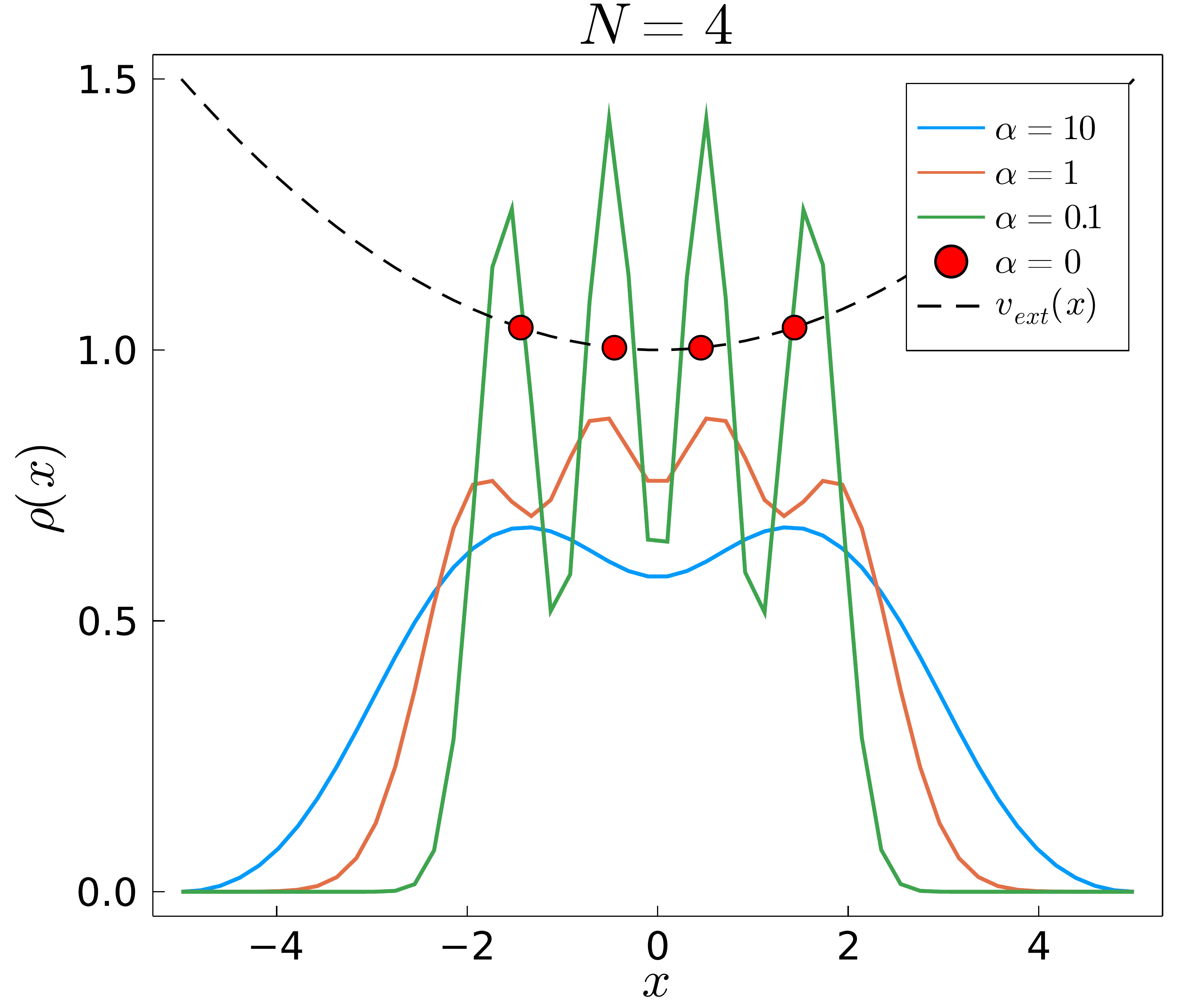}
\hskip 0.6cm
}
\subfigure[]
{
\includegraphics[width=7cm]{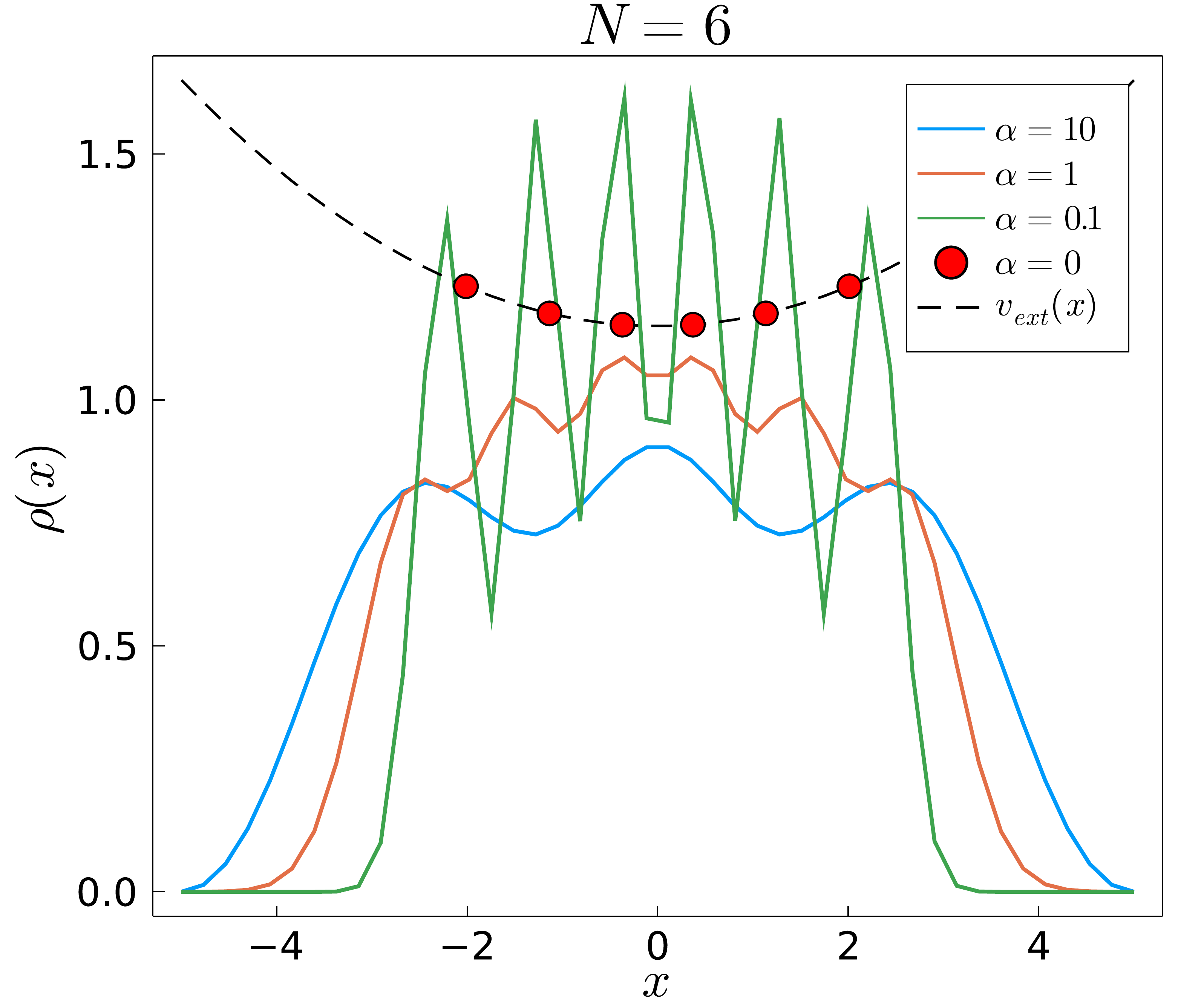}
}
\setlength{\abovecaptionskip}{-4pt}
\caption{The solid lines represent the single-electron densities (for 1D systems) with electron numbers $N=4$ and $N=6$.
The black dashed line represents the external potential.
The red solid balls represent the classical configurations that minimize \eqref{classic}.}
\label{fig:example1d:density}
\end{figure}

We then show the ground state single-electron densities (defined in \eqref{singledensity}) for systems with $N=4$ and 6 respectively in Figure \ref{fig:example1d:density}.
We observe from the pictures that as $\alpha$ decreases, the ground state densities depict a crossover from Fermi liquid to Wigner molecule.
For small $\alpha$, the electrons are concentrated on specific regions with $N$ peaks.
We compare the configurations obtained from the semi-classical limit (i.e. by Algorithm \ref{algorithm-ini}), and see that the classical configurations (the red solid balls) match with the peaks of electron densities very well.
This also explains why our construction of the initial state is good for systems with small $\alpha$, as it can capture the electron features in the strongly correlated regime.

\begin{figure}[htb!]
\centering
\subfigure[]
{
\includegraphics[width=4.5cm]{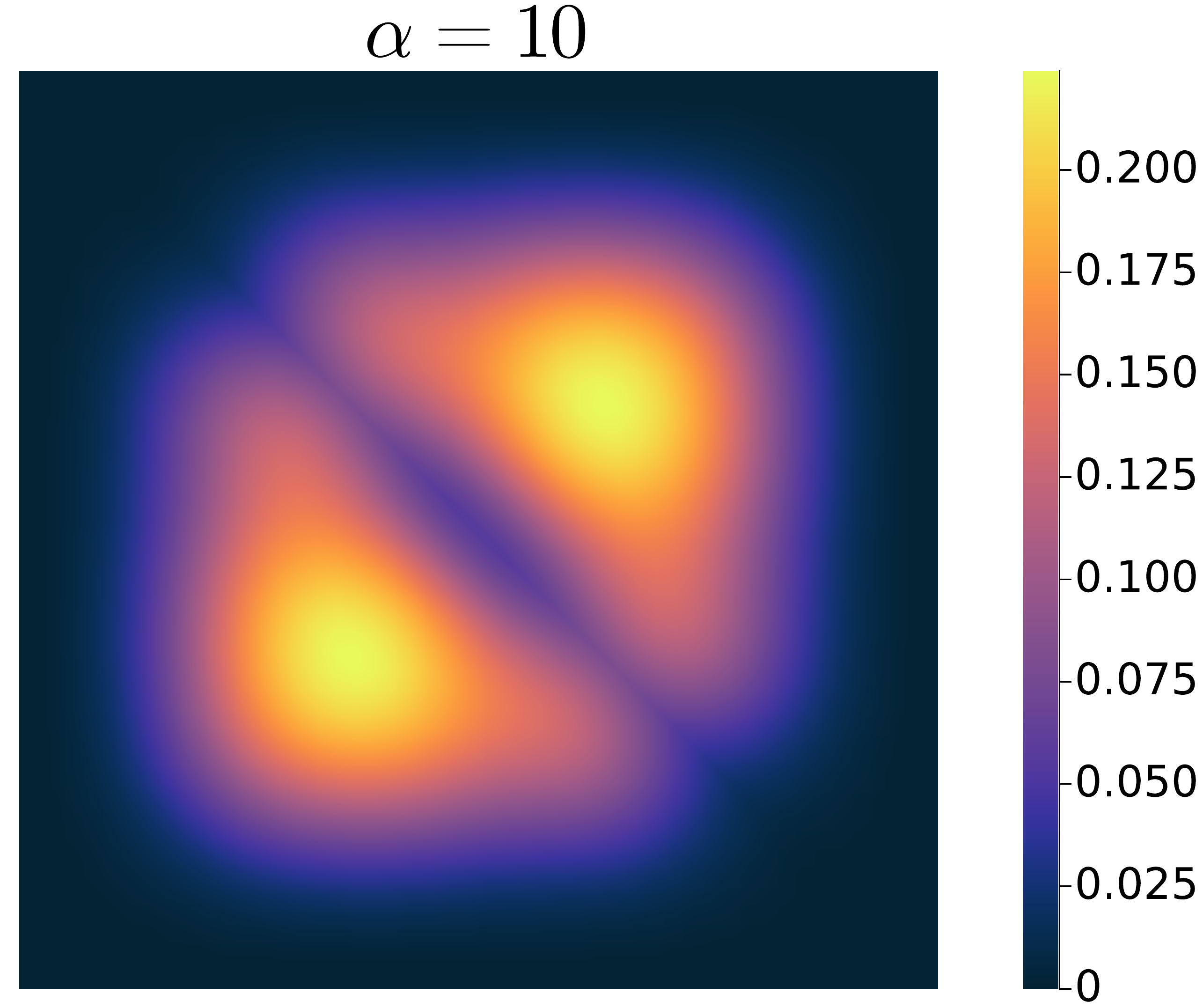}
}
\hskip 0.22cm
\subfigure[]
{
\includegraphics[width=4.5cm]{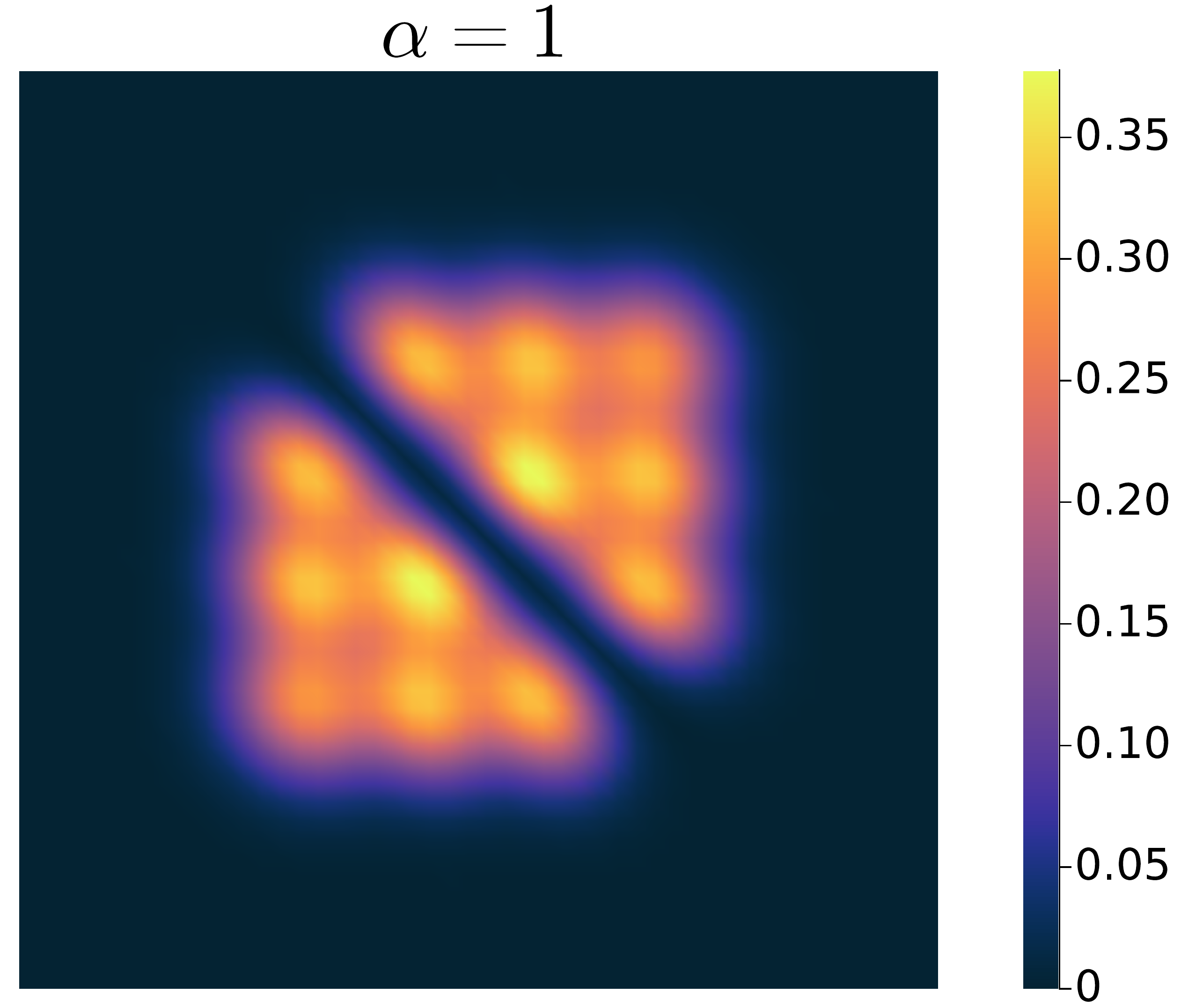}
\hskip 0.2cm
}
\subfigure[]
{
\includegraphics[width=4.5cm]{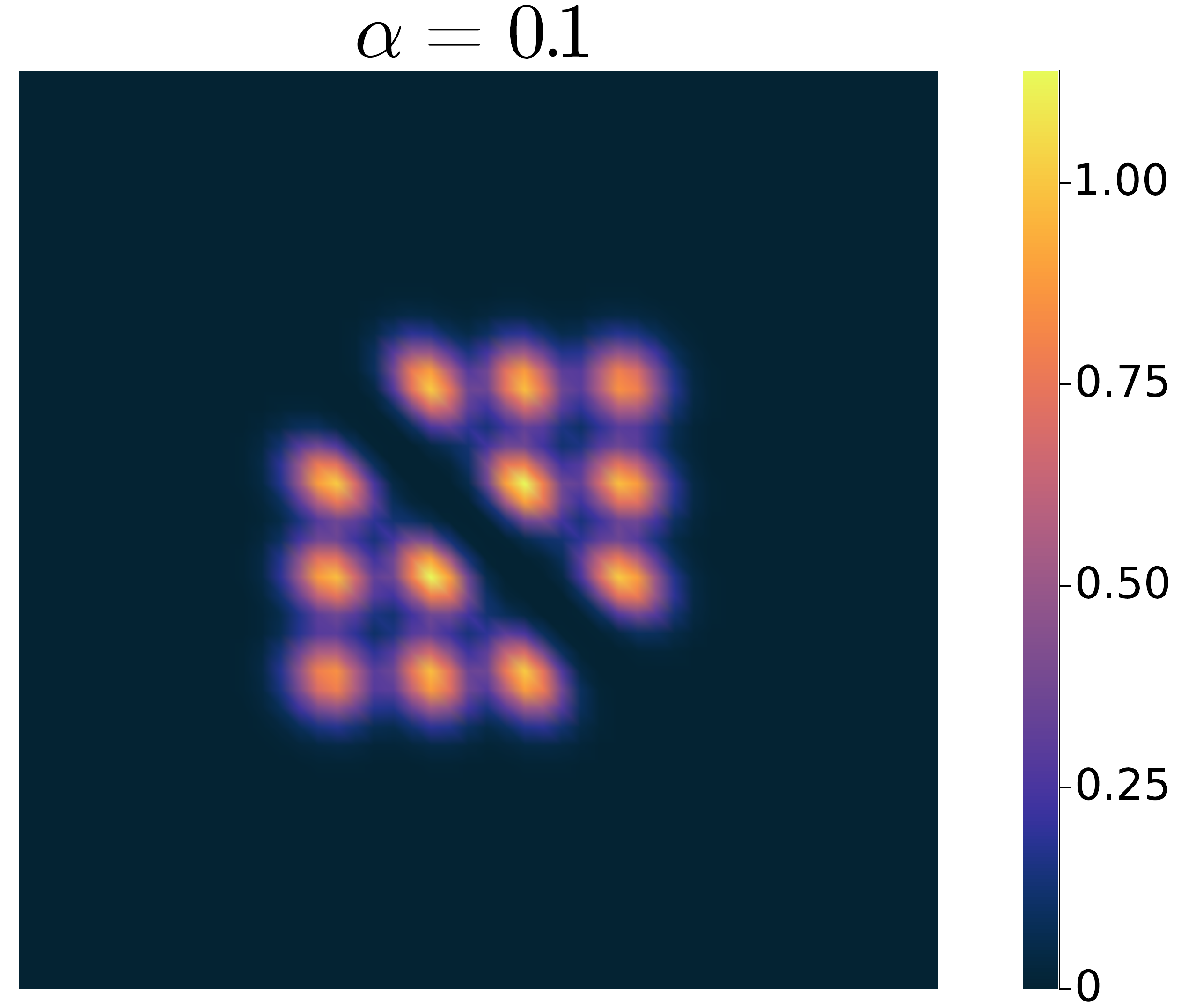}
}
\setlength{\abovecaptionskip}{-4pt}
\caption{Pair densities $\rho_2$ for 1D systems with electron number $N=4$.}
\label{fig:example1d:pairdensity}
\end{figure}

In order to visualize the internal ordering of the electrons of the wavefunction $\Psi$, we also plot the corresponding pair density distribution
\begin{eqnarray*}
\label{pairdensity}
\rho_2(x,x')= \binom{N}{2}\sum_{\st_1,\cdots,\st_N\in\Z_2}\int \big| \Psi(x,\st_1,x',\st_2,x_3,\st_3,\dots,x_N,\st_N) \big|^2 {\rm d}x_3\dots ~{\rm d}x_N .
\end{eqnarray*}
We show in Figure \ref{fig:example1d:pairdensity} the ground state pair densities for systems with electrons number $N=4$ and scaling parameters $\alpha=10,1,0.1$. 
We observe that the pair densities are always depleted near the diagonal $x=x'$, a phenomenon known as ``exchange holes". 
We also observe that when $\alpha$ is large, the pair density is smooth; when $\alpha$ is small, the pair density becomes localized and shows a clear ``correlation hole" in the Wigner regime.

\subsection{2D systems}
\label{sec:numerics2d}

We consider two-dimensional circularly-symmetric quantum dots with parabolic confinement on $\Omega=[-5,5]^2$ and systems with electron numbers $N=3,4,5,6$.
The external potential is given by $v_{\rm{ext}}(x,y) = \omega (x^2 + y^2)$ with the confinement strength $\omega$ between 0.1 and 2. 
We use a finite element mesh with $M_h = 30\times 30$ interior nodes for discretizaiton. 

\begin{figure}[htb!]
\centering
\subfigure[]
{
\includegraphics[width=4.5cm]{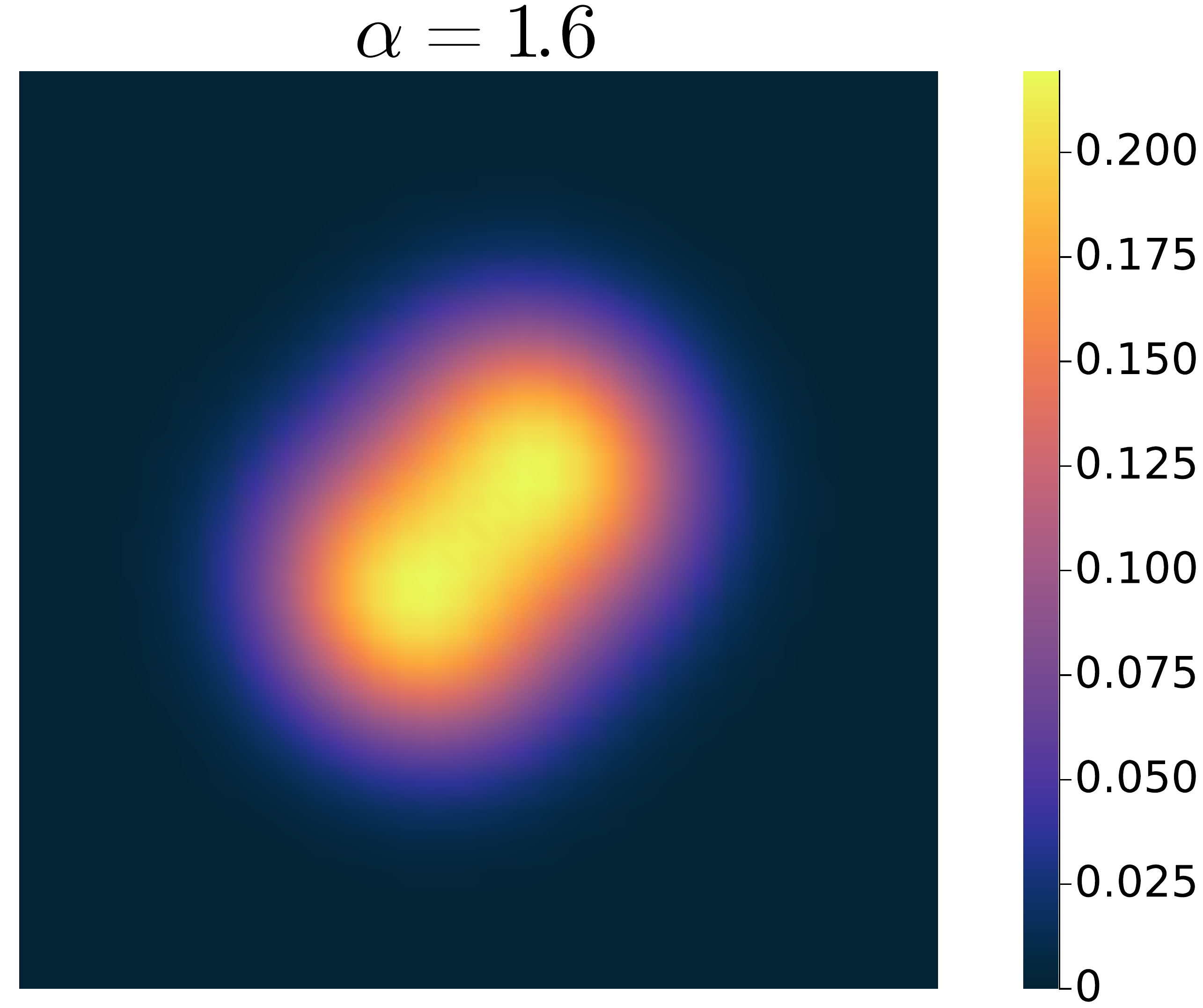}
\hskip 0.35cm
}
\subfigure[]
{
\includegraphics[width=4.5cm]{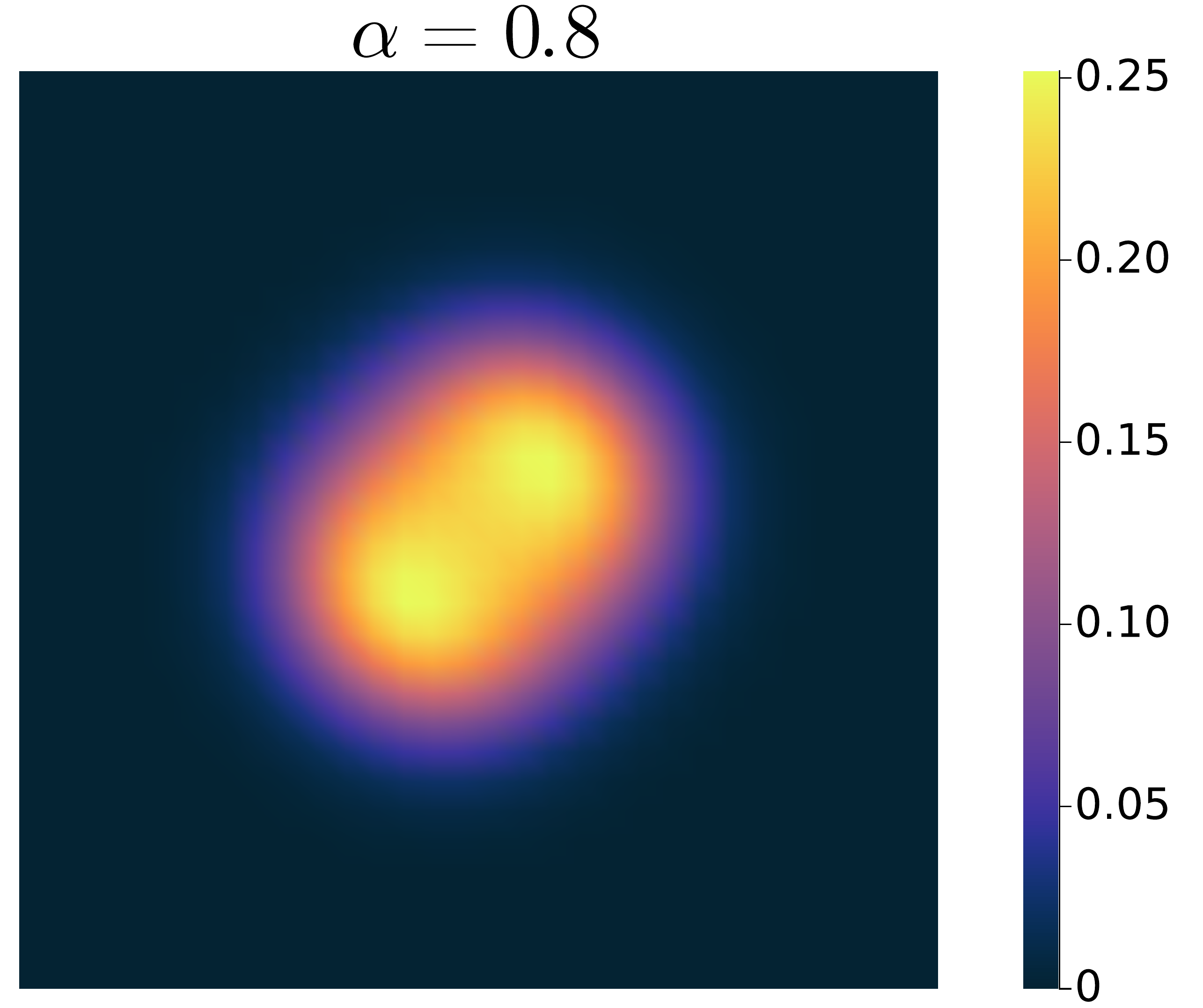}
}\\
\vskip 0.1cm
\hskip 0.1cm
\subfigure[]
{
\includegraphics[width=4.5cm]{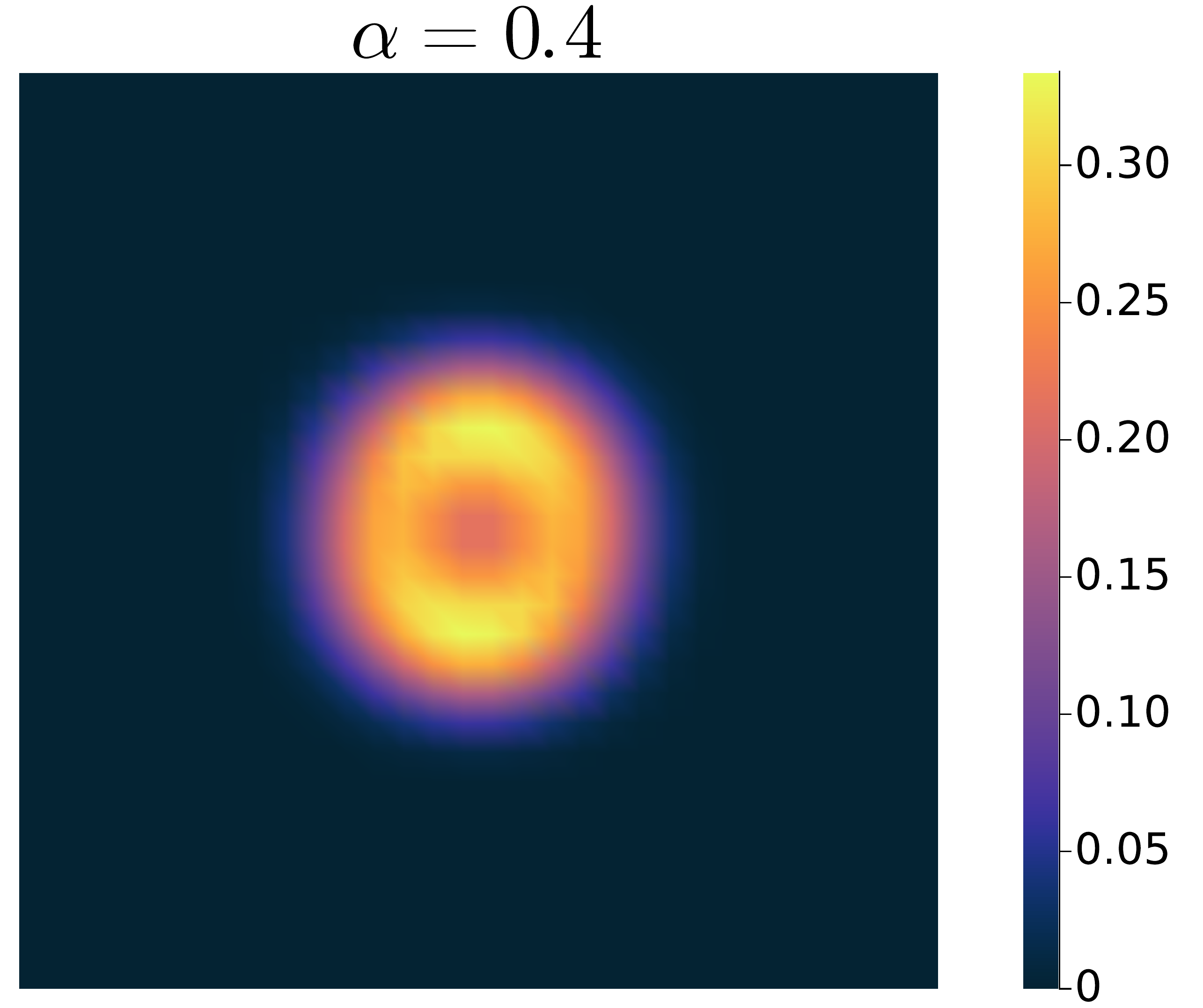}
\hskip 0.3cm
}
\subfigure[]
{
\includegraphics[width=4.5cm]{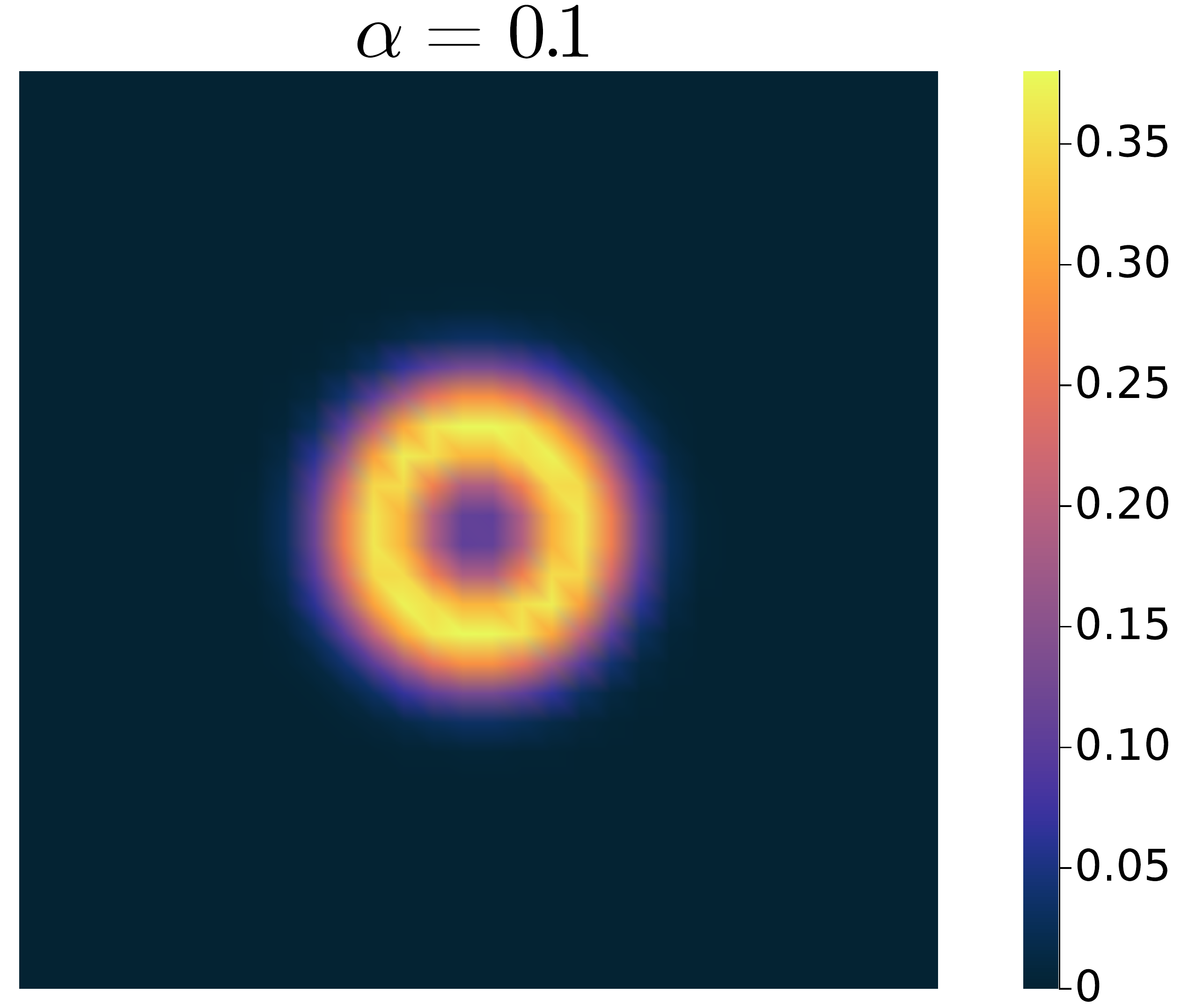}
}
\setlength{\abovecaptionskip}{-4pt}
\caption{Ground state densities with $N=3$.}
\label{fig:example2d:cross-over}
\end{figure}

We present the ground state densities of $N=3$, $\omega = 0.2$ in Figure \ref{fig:example2d:cross-over} with different scaling parameters $\alpha$. 
We observe a clear crossover from Fermi liquid structure to Wigner localization:
when $\alpha = 1.6$, the density is dumbbell shaped and concentrated on the middle area; when $\alpha = 0.8$, the density is still dumbbell shaped but there is a little dip in the center; when $\alpha = 0.4$, the density is still relatively smooth but becomes a thick ring; when $\alpha = 0.1$, the density becomes much sharper, forming a thin ring centered at the origin. 

\begin{figure}[htb!]
\centering
\subfigure[]
{
\includegraphics[width=4.5cm]{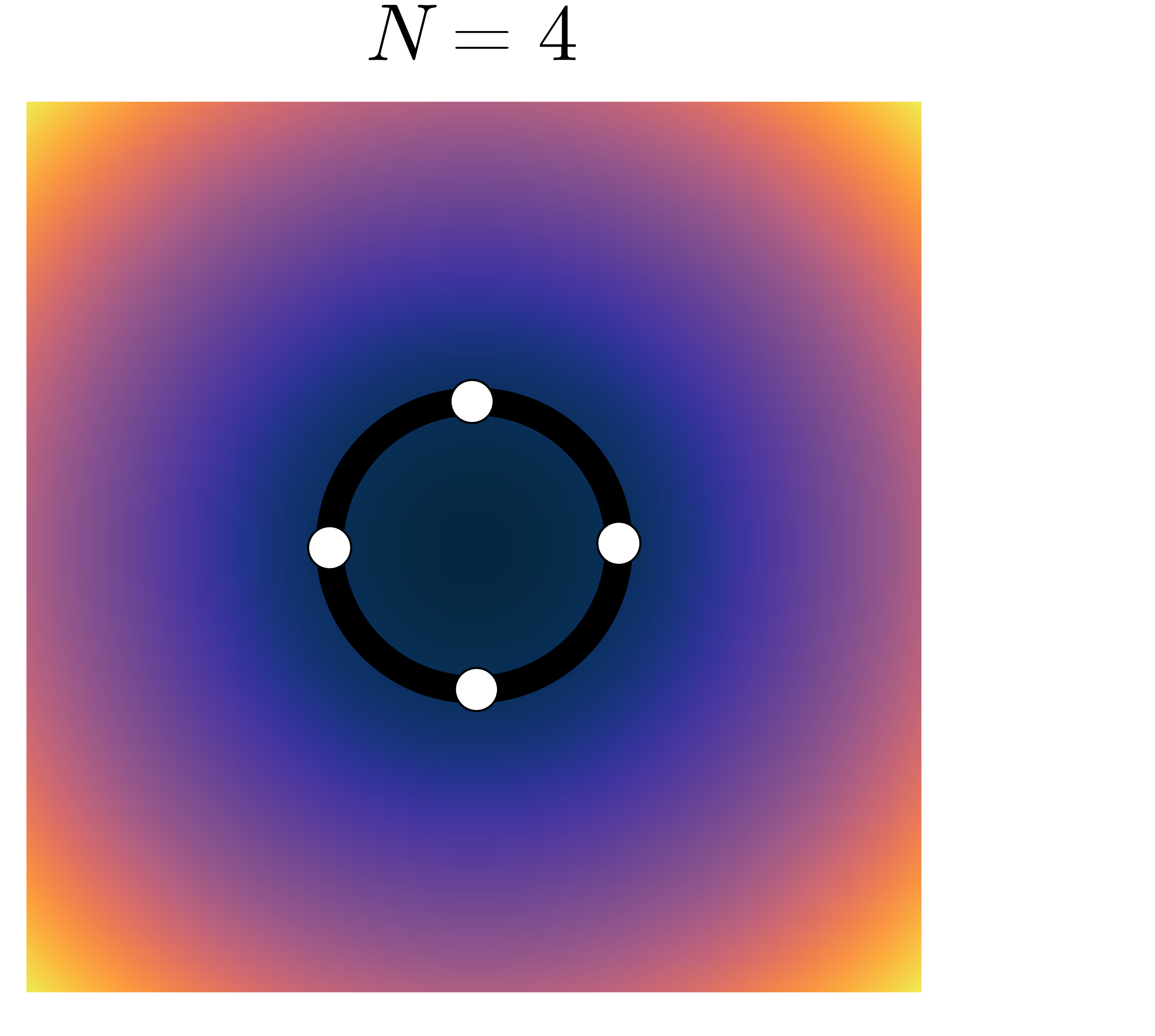}
}
\hskip 0.3cm
\subfigure[]
{
\includegraphics[width=4.5cm]{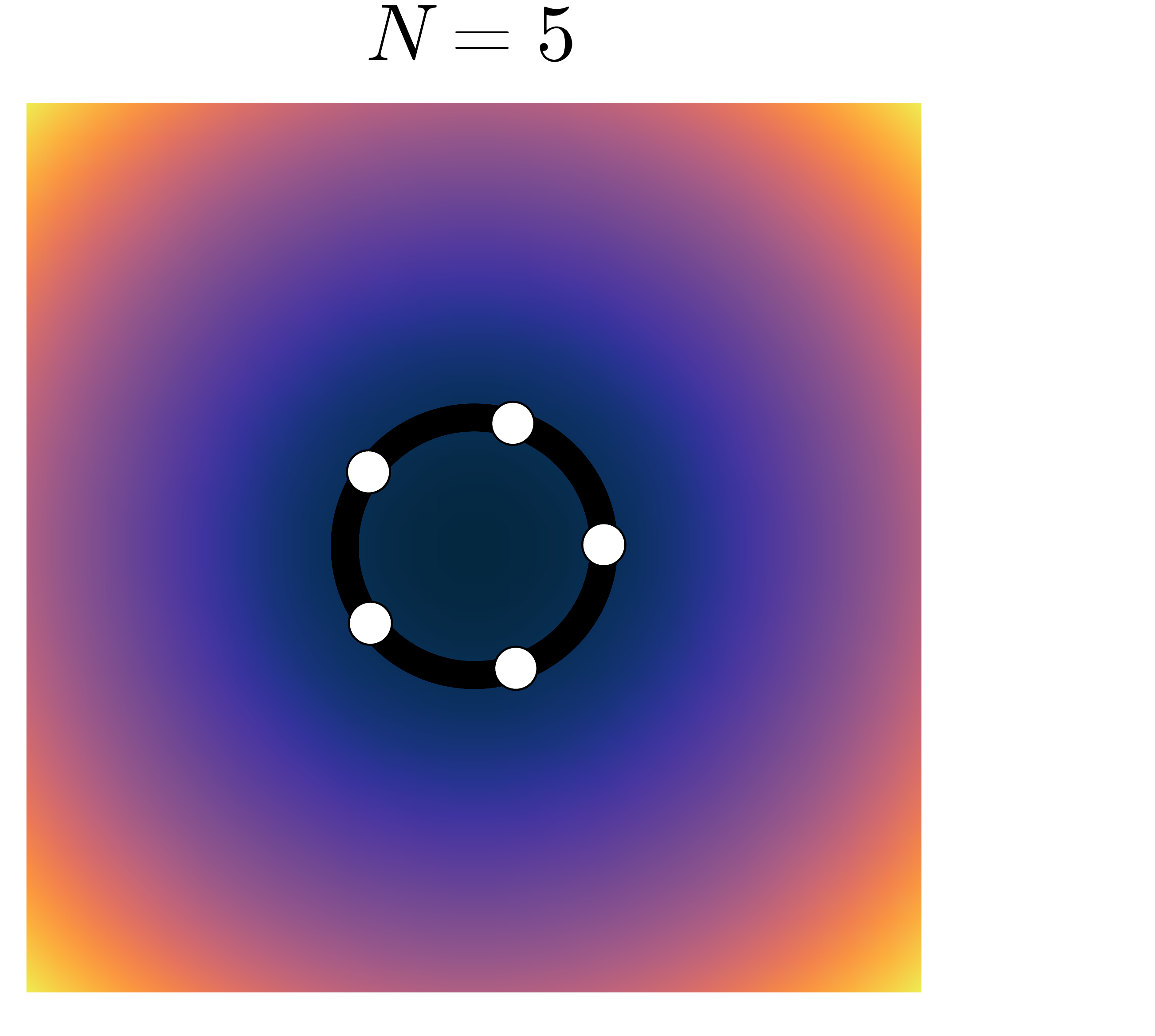}
}
\hskip 0.3cm
\subfigure[]
{
\includegraphics[width=4.5cm]{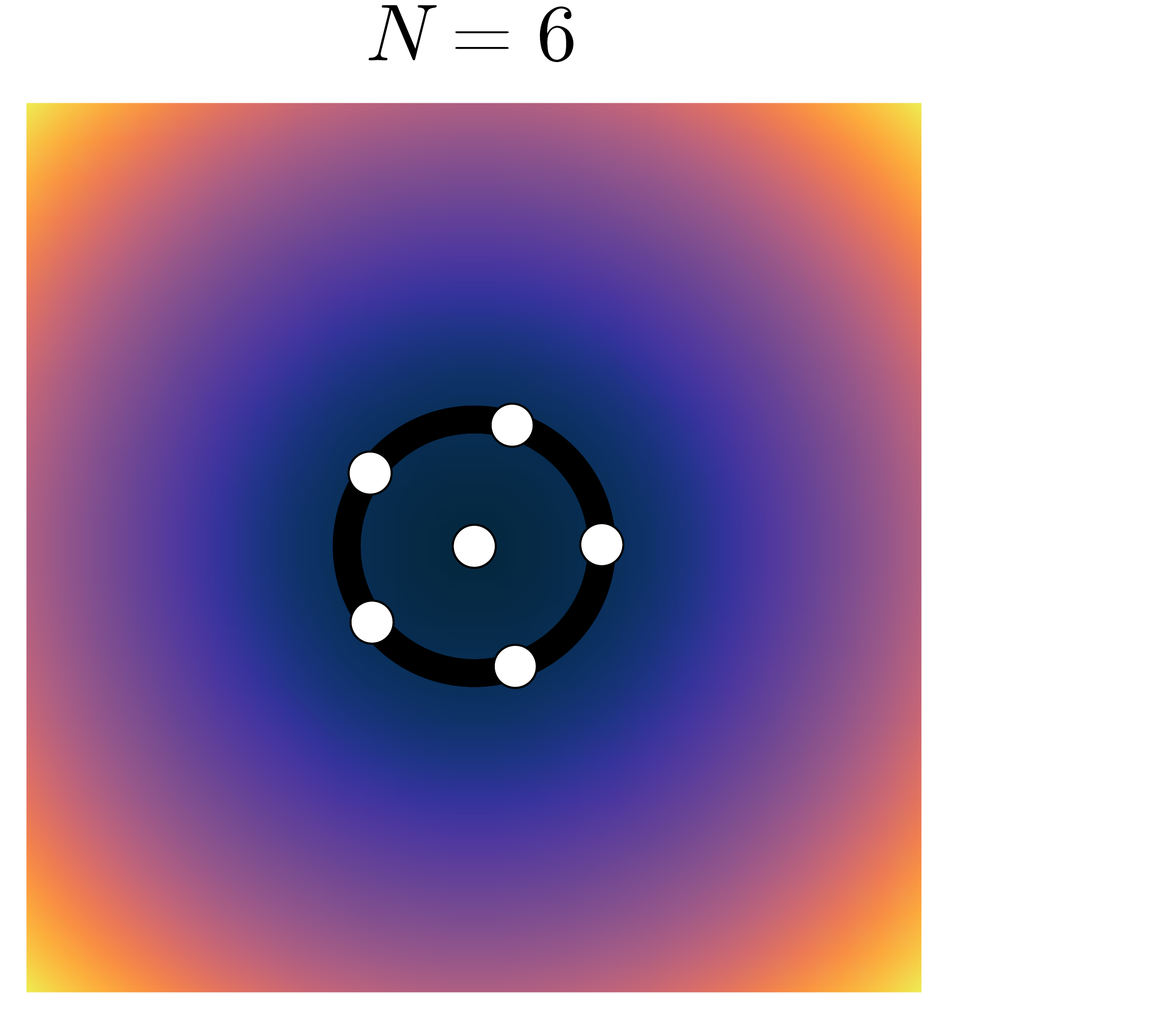}
}\\
\vskip -12pt
\hskip 0.1cm
\subfigure[]
{
\includegraphics[width=4.5cm]{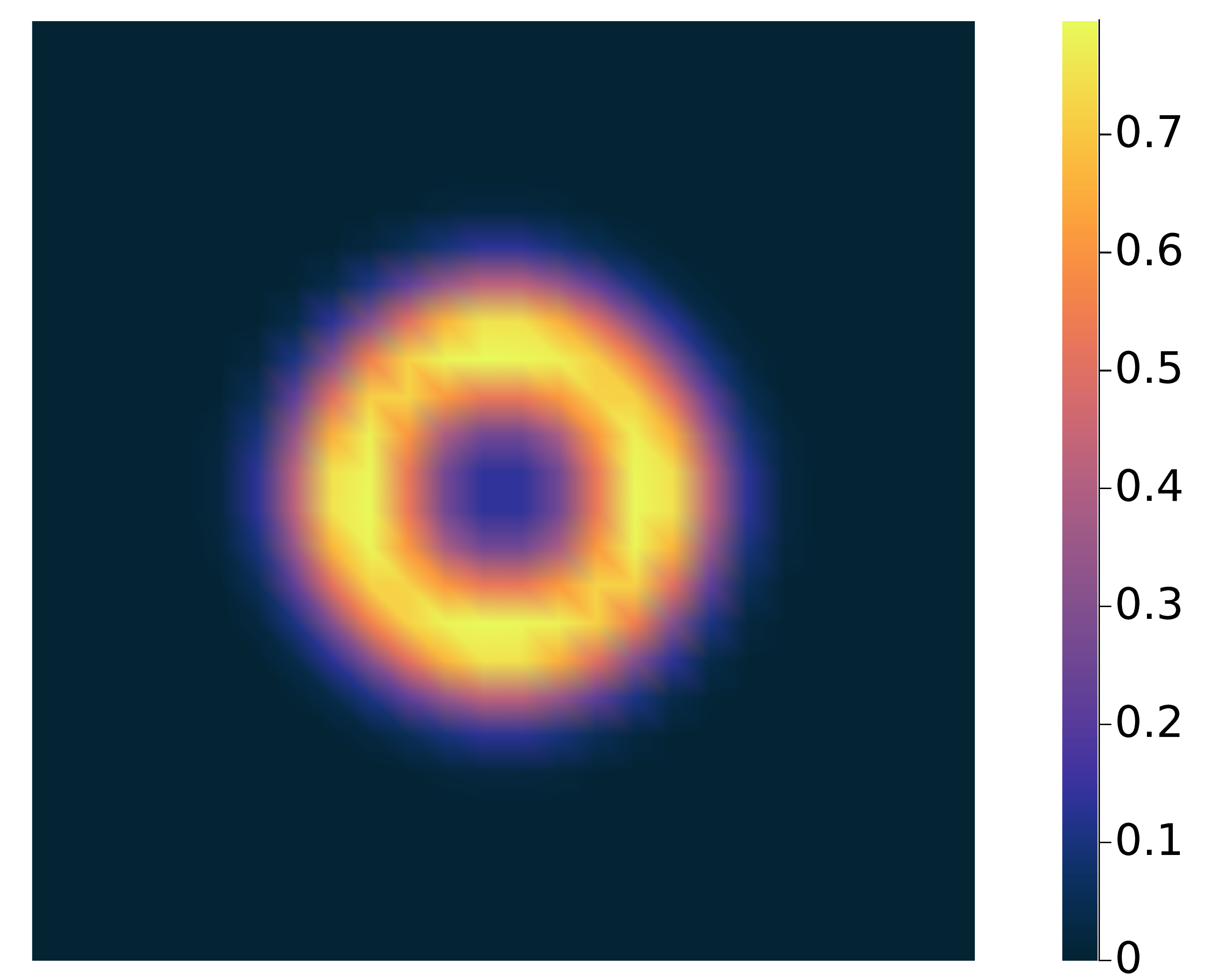}
}
\hskip 0.3cm
\subfigure[]
{
\includegraphics[width=4.5cm]{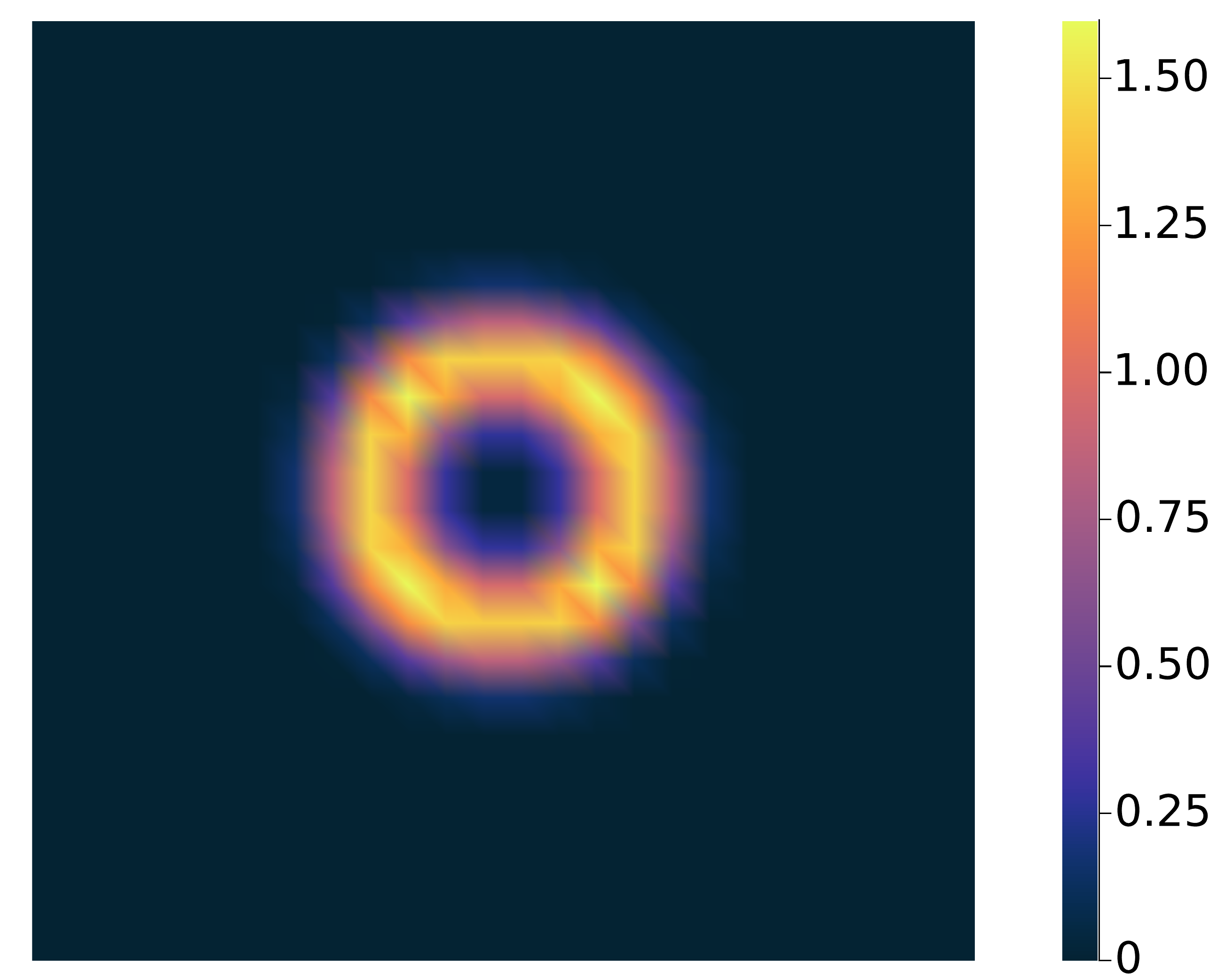}
}
\hskip 0.3cm
\subfigure[]
{
\includegraphics[width=4.5cm]{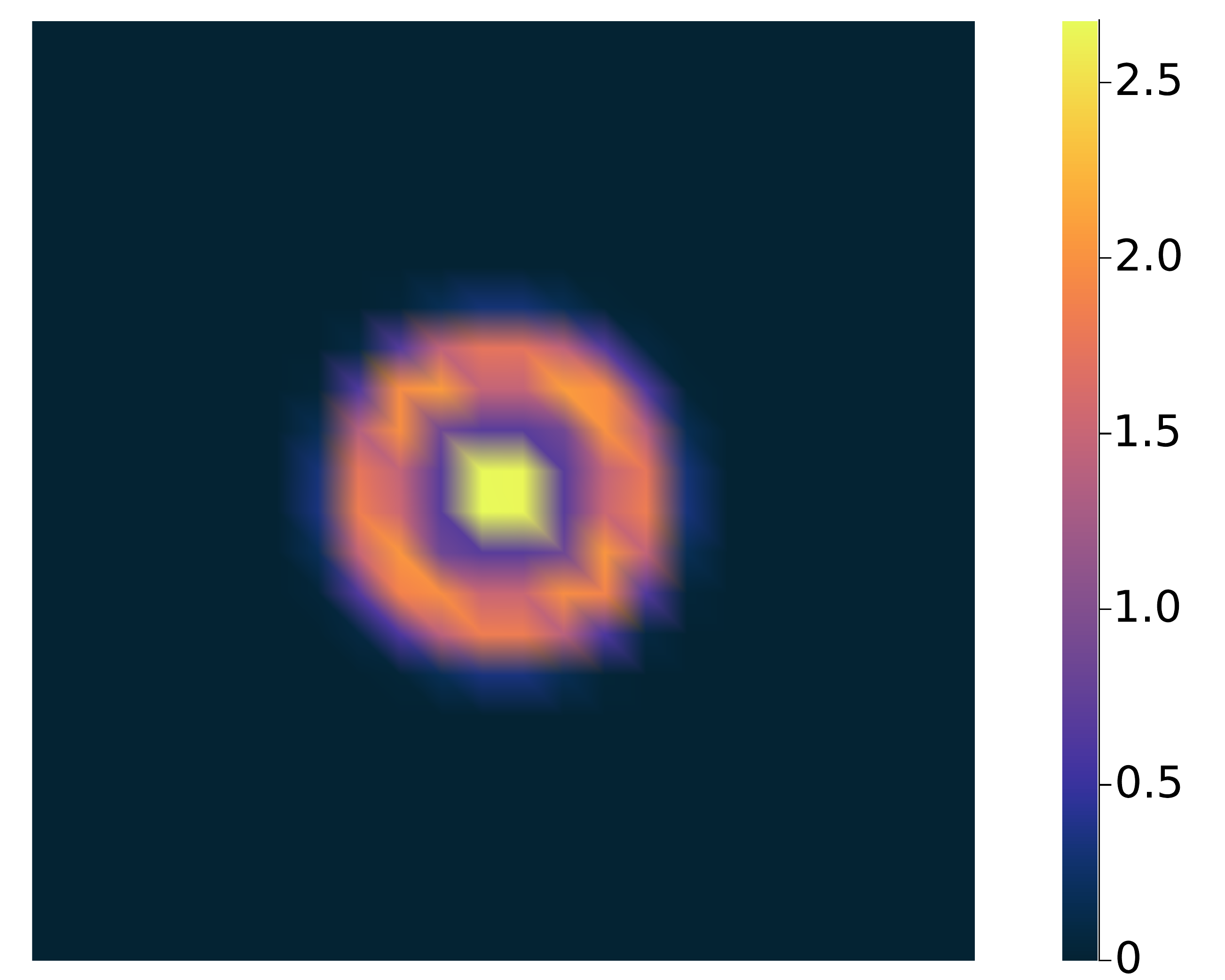}
}
\setlength{\abovecaptionskip}{-4pt}
\caption{The upper pictures show the external potentials (the background), one classical configuration that minimize \eqref{classic} (the white balls), and all possible such configurations (the black circles).
The lower pictures show the ground state densities (plot restricted on $[-3,3]^2$) with $N = 4, 5, 6$ and $\alpha = 0.1$.}
\label{fig:example2d:density}
\end{figure}

We further compare the Wigner localization for different particle numbers.
We take $\alpha=0.1$ and show in Figure \ref{fig:example2d:density} the external potentials, the classical configurations from Algorithm \ref{algorithm-ini}, and the ground state densities at the Wigner localization regime for $N=4,5$ and 6 respectively.
We observe that for $N=4$ and 5, the densities are concentrated on a ring, while for $N=6$ the density has one central electron (concentrated at the origin) with other electrons localized on a sharp ring surrounding it (which integrates to five particles).
This phase transition (with respect to the particle numbers) display the right classical filling for the spatial shells. 
We compare the configurations obtained from the semi-classical limit, which are consistent with the simulations and support that our construction of the initial state can capture the features of 2D strong correlation.
%
%

\section{Conclusions}
\label{sec:conclusion}
\setcounter{equation}{0}

In this work, we propose a configuration interaction algorithm for simulating the ground state of the Wigner localized systems.
The novelty of our algorithm lies in the combination of a finite elements discretization, a clever choice of the initial state designed particularly for the strongly correlated systems, and a selection of the determinants on the fly.
The algorithm can systematically resolve the sharp localization of the wavefunction and avoid exponential complexity of the many-body problem.

\subsection*{Acknowledgments}

The authors thank Genevi\`{e}ve Dusson for developing the finite elements based CI codes together and Gero Friesecke for inspiring conversations on the topic of Wigner localization.
This work was supported by the National Key R\&D Program of China (No. 2020YFA0712900). HC's work was also supported by National Natural Science Foundation of China (No. NSFC11971066).

\small
\bibliographystyle{plain}
\bibliography{main.bib}

\begin{thebibliography}{10}

\bibitem{Anderson18}
J.~S.~M. Anderson, F.~Heidar-Zadeh, and P.~W. Ayers.
\newblock Breaking the curse of dimension for the electronic {S}chr\"{o}dinger
  equation with functional analysis.
\newblock {\em Comput. Theo. Chem.}, 1142:66--77, 2018.

\bibitem{Andrei1988}
E.~Y. Andreia, G.~Deville, D.~C. Glattli, F.~I.~B. Williams, E.~Paris, and
  B.~Etienne.
\newblock Observation of a magnetically induced {W}igner solid.
\newblock {\em Phys. Rev. Lett.}, 60:2765--2768, 1988.

\bibitem{Ashcroft1976}
N.~W. Ashcroft and N.~D. Mermin.
\newblock {\em Solid State Physics}.
\newblock New York: Holt, Rinehart and Winston, 1976.

\bibitem{Auslaender05}
O.~M. Auslaender, H.~Steinberg, A.~Yacoby, Y.~Tserkovnyak, B.~I. Halperin,
  K.~W. Baldwin, L.~N. Pfeiffer, and K.~W. West.
\newblock Spin-charge separation and localization in one dimension.
\newblock {\em Science}, 308:88--92, 2005.

\bibitem{Ballester10}
A.~Ballester, J.~M. Escart\'in, J.~L. Movilla, M.~Pi, and J.~Planelles.
\newblock Mixed correlation phases in elongated quantum dots.
\newblock {\em Phys. Rev. B}, 82:115405, 2010.

\bibitem{Bao2012}
G.~Bao, G.~Hu, and D.~Liu.
\newblock An h-adaptive finite element solver for the calculations of the
  electronic structures.
\newblock {\em J. Comput. Phys.}, 231:4967--4979, 2012.

\bibitem{bednarek03}
S.~Bednarek, B.~Szafran, T.~Chwiej, and J.~Adamowski.
\newblock Effective interaction for charge carriers confined in
  quasi-one-dimensional nanostructures.
\newblock {\em Phys. Rev. B}, 68(4), 2003.

\bibitem{Julia}
J.~Bezanson, A.~Edelman, S.~Karpinski, and V.B. Shah.
\newblock Julia: {A} fresh approach to numerical computing.
\newblock {\em SIAM Review}, 59:65--98, 2017.

\bibitem{Booth2009}
G.~H. Booth, A.~J.~W. Thom, and A.~Alavi.
\newblock Fermion {Monte Carlo} without fixed nodes: A game of life, death, and
  annihilation in $\mathrm{S}$later determinant space.
\newblock {\em J. Chem. Phys.}, 131:054106, 2009.

\bibitem{Buenker1974}
R.~J. Buenker and S.~D. Peyerimhoff.
\newblock Individualized configuration selection in {CI} calculations with
  subsequent energy extrapolation.
\newblock {\em Theor. Chim. Acta}, 35:1974, 33-58.

\bibitem{Buhmann1991}
H.~Buhmann, W.~Joss, K.~V. Klitzing, I.~V. Kukushkin, A.~S. Plaut, G.~Martinez,
  K.~Ploog, and V.~B. Timofeev.
\newblock Novel magneto-optical behavior in the {W}igner-solid regime.
\newblock {\em Phys. Rev. Lett.}, 66:926--929, 1991.

\bibitem{Buttazzo2012}
G.~Buttazzo, L.~Pascale, and P.~Gori-Giorgi.
\newblock Optimal-transport formulation of electronic density-functional
  theory.
\newblock {\em Phys. Rev. A}, 85:062502, 2012.

\bibitem{Chen2014}
H.~Chen, X.~Dai, X.~Gong, L.~He, and A.~Zhou.
\newblock Adaptive finite element approximations for {Kohn-Sham} models.
\newblock {\em Multiscale Model. Simul.}, 12:1828--1869, 2014.

\bibitem{Chen2015}
H.~Chen and G.~Friesecke.
\newblock Pair densities in density functional theory.
\newblock {\em Multiscale Model. Simul.}, 13:1259--1289, 2015.

\bibitem{Chen2003}
Y.~Chen, R.~M. Lewis, L.~W. Engel, D.~C. Tsui, P.~D. Ye, L.~N. Pfeiffer, and
  K.~W. West.
\newblock Microwave resonance of the {2D Wigner} crystal around integer
  $\mathrm{L}$andau fillings.
\newblock {\em Phys. Rev. Lett.}, 91:016801, 2003.

\bibitem{Cotar2013}
C.~Cotar, G.~Friesecke, and C.~Kl\"{u}ppelberg.
\newblock Density functional theory and optimal transportation with {C}oulomb
  cost.
\newblock {\em Commun. Pure Appl. Math.}, 66:548--599, 2013.

\bibitem{friesecke15}
C.~Cotar, G.~Friesecke, and B.~Pass.
\newblock Infinite-body optimal transport with {C}oulomb cost.
\newblock {\em Calc. Var.}, 54:717--742, 2015.

\bibitem{Deng2016}
H.~Deng, Y.~Liu, I.~Jo, L.~N. Pfeiffer, K.~W. West, K.~W. Baldwin, and
  M.~Shayegan.
\newblock Commensurability oscillations of composite fermions induced by the
  periodic potential of a {W}igner crystal.
\newblock {\em Phys. Rev. Lett.}, 117:096601, 2016.

\bibitem{Deshpande08}
V.~V. Deshpande and M.~Bockrath.
\newblock The one-dimensional {W}igner crystal in carbon nanotubes.
\newblock {\em Nat. Phys.}, 4:314, 2008.

\bibitem{Dirac1939}
P.~A.~M. Dirac.
\newblock A new notation for quantum mechanics.
\newblock {\em Math. Proc. Camb. Philos. Soc.}, 35:416--418, 1939.

\bibitem{Xue}
G.~Dusson and X.~Quan.
\newblock \url{https://github.com/dussong/PairDensities.jl}, 2022.

\bibitem{fournais18}
S.~N Fournais, M.~Lewin, and J.~P. Solovej.
\newblock The semi-classical limit of large fermionic systems.
\newblock {\em Calc. Var.}, 57:105, 2018.

\bibitem{friesecke22}
G.~Friesecke, A.~Gerolin, and P.~Gori-Giorgi.
\newblock The strong-interaction limit of density functional theory.
\newblock arXiv:2202.09760, 2022.

\bibitem{friesecke21}
G.~Friesecke, A.~S. Schulz, and D.~V\"{o}gler.
\newblock Genetic column generation: {F}ast computation of high-dimensional
  multi-marginal optimal transport problems.
\newblock {\em SIAM J. Sci. Comput.}, 44:A1632--A1654, 2022.

\bibitem{Gavini2007}
V.~Gavini, J.~Knap, K.~Bhattacharya, and M.~Ortiz.
\newblock Non-periodic finite-element formulation of orbital-free density
  functional theory.
\newblock {\em J. Mech. Phys. Solids}, 55:669--696, 2007.

\bibitem{Ghosal06}
A.~Ghosal, A.~D. Guclu, C.~J. Umrigar, D.~Ullmo, and H.~U. Baranger.
\newblock Correlation-induced inhomogeneity in circular quantum dots.
\newblock {\em Nature Phys.}, 2:336--340, 2006.

\bibitem{Goldys1992}
E.~M. Goldys, S.~A. Brown, R.~B. Dunford, A.~G. Davies, R.~Newbury, R.~G.
  Clark, P.~E. Simmonds, J.~J. Harris, and C.~T. Foxon.
\newblock Magneto-optical probe of two-dimensional electron liquid and solid
  phases.
\newblock {\em Phys. Rev. B}, 46:7957--7960, 1992.

\bibitem{Giorgi2009}
P.~Gori-Giorgi, M.~Seidl, and G.~Vignale.
\newblock Density functional theory for strongly interacting electrons.
\newblock {\em Phys. Rev. Lett.}, 103:166402, 2009.

\bibitem{Greene2019}
S.~M. Greene, R.~J. Webber, J.~Weare, and T.~C. Berkelbach.
\newblock Beyond walkers in stochastic quantum chemistry: {R}educing error
  using fast randomized iteration.
\newblock {\em J. Chem. Theory Comput.}, 15:4834--4850, 2019.

\bibitem{Grimes1979}
C.~C. Grimes and G.~Adams.
\newblock Evidence for a liquid-to-crystal phase transition in a classical,
  two-dimensional sheet of electrons.
\newblock {\em Phys. Rev. Lett.}, 42:795--798, 1979.

\bibitem{Helgaker2000}
T.~Helgaker, P.~J\o rgensen, and J.~Olsen.
\newblock {\em Molecular Electronic Structure Theory}.
\newblock John Wiley \& Sons Ltd, 2000.

\bibitem{Holmes2016}
A.~A. Holmes, N.~M. Tubman, and C.~J. Umrigar.
\newblock Heat-bath configuration interaction: {A}n efficient selected {CI}
  algorithm inspired by heat-bath sampling.
\newblock {\em J. Chem. Theory Comput.}, 12:3674--3680, 2016.

\bibitem{Huron1973}
B.~Huron, J.~P. Malrieu, and P.~Rancurel.
\newblock Iterative perturbation calculations of ground and excited state
  energies from multiconfigurational zeroth-order wavefunctions.
\newblock {\em J. Chem. Phys.}, 58:5745--5759, 1973.

\bibitem{Konik2002}
R.~M. Konik and P.~Fendley.
\newblock Haldane-gapped spin chains as {L}uttinger liquids: {C}orrelation
  functions at finite field.
\newblock {\em Phys. Rev. B}, 66:144416, 2002.

\bibitem{Li2016}
L.~Li, T.~E. Baker, S.~R. White, and K.~Burke.
\newblock Pure density functional for strong correlation and the thermodynamic
  limit from machine learning.
\newblock {\em Phys. Rev. B}, 94:245129, 2016.

\bibitem{Li2019}
Y.~Li, J.~Lu, and Z.~Wang.
\newblock Coordinate-wise descent methods for leading eigenvalue problem.
\newblock {\em SIAM J. Sci. Comput.}, 41:A2681–A2716, 2019.

\bibitem{Lim2017}
L.~H. Lim and J.~Weare.
\newblock Fast randomized iteration: {Diffusion Monte Carlo} through the lens
  of numerical linear algebra.
\newblock {\em SIAM. Rev.}, 59:547--587, 2017.

\bibitem{Lu2020}
J.~Lu and Z.~Wang.
\newblock The full configuration interaction quantum {Monte Carlo} method in
  the lens of inexact power iteration.
\newblock {\em SIAM J. Sci. Comput.}, 42(1):B1--B29, 2020.

\bibitem{Malet2012}
F.~Malet and P.~Gori-Giorgi.
\newblock Strong correlation in {Kohn-Sham} density functional theory.
\newblock {\em Phys. Rev. Lett.}, 109:246402, 2012.

\bibitem{Malet2013}
F.~Malet, A.~Mirtschink, J.~C. Cremon, S.~M. Reimann, and P.~Gori-Giorgi.
\newblock {Kohn-Sham} density functional theory for quantum wires in arbitrary
  correlation regimes.
\newblock {\em Phys. Rev. B}, 87:115146, 2013.

\bibitem{Mendez1983}
E.~E. Mendez, M.~Heiblum, L.~L. Chang, and L.~Esaki.
\newblock High-magnetic-field transport in a dilute two-dimensional electron
  gas.
\newblock {\em Phys. Rev. B}, 28:4886--4888, 1983.

\bibitem{mendl14}
C.~B. Mendle, F.~Malet, and P.~Gori-Giorgi.
\newblock Wigner localization in quantum dots from {Kohn-Sham} density
  functional theory without symmetry breaking.
\newblock {\em Phys. Rev. B}, 89:125106, 2014.

\bibitem{Pask2001}
J.~E. Pask, B.~M. Klein, P.~A. Sterne, and C.~Y. Fong.
\newblock Finite-element methods in electronic-structure theory.
\newblock {\em Comput. Phys. Comm.}, 135:1--34, 2001.

\bibitem{Postma2001}
H.~W.~C. Postma, T.~Teepen, Z.~Yao, M.~Grifoni, and C.~Dekker.
\newblock Carbon nanotube single-electron transistors at room temperature.
\newblock {\em Science}, 293:76--79, 2001.

\bibitem{Seidl2007}
M.~Seidl, P.~Gori-Giorgi, and A.~Savin.
\newblock Strictly correlated electrons in density functional theory:
  $\mathrm{A}$ general formulation with applications to spherical densities.
\newblock {\em Phys. Rev. A}, 75:042511, 2007.

\bibitem{seidl1999}
M.~Seidl, J.~P. Perdew, and M.~Levy.
\newblock Strictly correlated electrons in density-functional theory.
\newblock {\em Phys. Rev. A}, 59:51, 1999.

\bibitem{serfaty18}
S.~Serfaty.
\newblock Systems of points with {C}oulomb interactions.
\newblock {\em ICM}, 2018:935--977, 2019.

\bibitem{Sharma2017}
S.~Sharma, A.~A. Holmes, G.~Jeanmairet, A.~Alavi, and C.~J. Umrigar.
\newblock Semistochastic heat-bath configuration interaction method:
  $\mathrm{S}$elected configuration interaction with semistochastic
  perturbation theory.
\newblock {\em J. Chem. Theory Comput.}, 13:1595–1604, 2017.

\bibitem{Suryanarayana2010}
P.~Suryanarayana, V.~Gavini, T.~Blesgen, K.~Bhattacharya, and M.~Ortiz.
\newblock Non-periodic finite-element formulation of {K}ohn–{S}ham density
  functional theory.
\newblock {\em J. Mech. Phys. Solids}, 58:256--280, 2010.

\bibitem{Taylor08}
J.~M. Taylor and T.~Calarco.
\newblock Wigner crystals of ions as quantum hard drives.
\newblock {\em Phys. Rev. A}, 78:062331, 2008.

\bibitem{Tsuchida1996}
E.~Tsuchida and M.~Tsukada.
\newblock Adaptive finite-element method for electronic-structure calculations.
\newblock {\em Phys. Rev. B}, 54:7602, 1996.

\bibitem{Tubman2016}
N.~M. Tubman, J.~Lee, T.~Y. Takeshita, M.~Head-Gordon, and K.~B. Whaley.
\newblock A deterministic alternative to the full configuration interaction
  quantum $\mathrm{M}$onte $\mathrm{C}$arlo method.
\newblock {\em J. Chem. Phys.}, 145:044112, 2016.

\bibitem{Lu2019}
Z.~Wang, Y.~Li, and J.~Lu.
\newblock Coordinate descent full configuration interaction.
\newblock {\em J. Chem. Theory Comput.}, 15(6):3558--3569, 2019.

\bibitem{Weiss06}
S.~Weiss, M.~Thorwart, and R.~Egger.
\newblock Charge qubit entanglement in double quantum dots.
\newblock {\em Europhys. Lett.}, 76:905, 2006.

\bibitem{Yannouleas07}
C.~Yannouleas and U.~Landman.
\newblock Symmetry breaking and quantum correlations in finite systems:
  {S}tudies of quantum dots and ultracold {B}ose gases and related nuclear and
  chemical methods.
\newblock {\em Rep. Prog. Phys.}, 70:2067, 2007.

\end{thebibliography}

\end{document}